%% file: falcon_main.tex
\documentclass[letterpaper]{article}
\usepackage[affil-it]{authblk}

\usepackage{amsmath}
\usepackage{amsthm}

\input{\commonDir preambleCommon}      %
\usepackage[numbers,sort&compress]{natbib}
\graphicspath{{./falcon/figures/}}

\usepackage{url}
\usepackage{fullpage}
\usepackage{setspace}

\makeatletter
\@addtoreset{algorithm}{section}
\makeatother

\title{A robust and efficient method for estimating enzyme
complex abundance and metabolic flux from expression data}

\date{\today}

\begin{document}

\author[1]{Brandon E. Barker%
  \thanks{Electronic address: \texttt{brandon.barker@cornell.edu}; Corresponding author}}

\newcounter{NarayanYipingCoAuthor}
\author[2]{Narayanan Sadagopan\footnote{contributed equally}%
\protect\setcounter{NarayanYipingCoAuthor}{\value{footnote}}%
}
\newcommand\NarayanYipingCoAuthorMark{\footnotemark[\value{NarayanYipingCoAuthor}]}%

\author[3]{Yiping Wang\protect\NarayanYipingCoAuthorMark}

\author[4,5]{Kieran Smallbone}

\author[3,6,7]{Christopher R. Myers}

\author[8]{Hongwei Xi}

\author[3,9]{Jason W. Locasale%
  \thanks{Electronic address: \texttt{locasale@cornell.edu}; Corresponding author}}

\author[3,9]{Zhenglong Gu%
  \thanks{Electronic address: \texttt{zg27@cornell.edu}; Corresponding author}}

\affil[1]{Center for Advanced Computing,
  Cornell University, 534 Rhodes Hall, Ithaca, NY, USA.}
\affil[2]{Program of Bioinformatics and Integrative Biology,
  University of Massachusetts Medical School,
  55 Lake Avenue North, Worcester, MA, USA.}
\affil[3]{Tri-Institutional Training Program in Computational
  Biology and Medicine, 1300 York Avenue, Box 194, New York, NY, USA.}
\affil[4]{School of Computer Science, The University of Manchester,
  Manchester, UK.}
\affil[5]{Manchester Center for Integrative Systems Biology, 
  The University of Manchester, \mbox{Manchester, UK.}}
\affil[6]{Laboratory of Atomic and Solid State Physics, Cornell
University, Ithaca, NY, USA.}
\affil[7]{Institute of Biotechnology, Cornell University, Ithaca, NY,
USA.}
\affil[8]{Department of Computer Science, Boston University,
  111 Cummington Street, Boston, MA, USA.}
\affil[9]{Division of Nutritional Sciences, Cornell University,
  Savage Hall, Ithaca, NY, USA.}

\newboolean{thesisStyle}               %
\setboolean{thesisStyle}{true}         %

\maketitle

\input{\commonDir documentHeadCommon}  %
                                       %

\begin{abstract}
\falconAbstractMotivation
\falconAbstractResults
\falconAbstractAvail
\end{abstract}

\def\suppOrApp{}

\input{\commonDir falcon}  %

\input{\commonDir falcon_appendix}   %

\bibliography{library}
\end{document}

%% file: common/preambleCommon.tex
%

\usepackage{caption}

\usepackage{algorithm}
\usepackage{algpseudocode}
\usepackage{float}
\usepackage{ifthen}
\usepackage{soul} 
\usepackage{tikz}
\usepackage[export]{adjustbox}
\usepackage{placeins}

\usepackage{fixltx2e}
\usepackage{upquote,textcomp}

\usepackage{palatino}
\usepackage{subcaption}
\usepackage{txfonts}
\usepackage{array}
\usepackage{calc}
\usepackage{pstricks}
\usepackage{graphicx}
\usepackage{epstopdf}
\usepackage{colortbl}
\usepackage{units}
\usepackage[makeroom]{cancel}

\usetikzlibrary{shapes, arrows}

\makeatletter

\newcommand\floatc@plainthin[2]{
\setbox\@tempboxa\hbox{{\@fs@cfont #1:} #2}
   \hbox to\hsize{\hfil\box\@tempboxa\hfil}\fi}
\newcommand\fs@plainthin{
  \def\@fs@cfont{\rmfamily}\let\@fs@capt\floatc@plainthin
  \def\@fs@pre{}
  \def\@fs@post{\vspace{-2.5em}}
  \def\@fs@mid{\vspace\abovecaptionskip\relax}
  \let\@fs@iftopcapt\iffalse}
\makeatother
\floatstyle{plainthin}
\newfloat{AlgFloat}{h}{lop}

\algnewcommand\algorithmicinput{\textbf{INPUT:}}
\algnewcommand\INPUT{\item[\algorithmicinput]}
\algnewcommand\algorithmicoutput{\textbf{OUTPUT:}}
\algnewcommand\OUTPUT{\item[\algorithmicoutput]}

\makeatletter
\@ifclassloaded{beamer}
{}
{

\newtheoremstyle{break}  
  {\topsep}   
  {\topsep}   
  {\itshape}  
  {-1em}      
  {\bfseries} 
  {}         
  {0pt}  
  {}          
\theoremstyle{break}

}
\makeatother

\newcommand{\E}[1]{\operatorname{E}\left(#1\right)}
\newcommand{\sgn}[1]{\operatorname{sgn}\left(#1\right)}

%% file: common/documentHeadCommon.tex
\newcommand{\captionpage}[2]{
\newpage
\vspace*{\fill}
\begin{center}
\captionof{#1}{#2}
\end{center}
\vspace{\fill}
\newpage
}

\tikzstyle{line} = [draw, very thick, color=black!50, -latex']
\tikzstyle{cloud} = [draw, ellipse,fill=red!20, node distance=2em]
\tikzstyle{decision} = [diamond, draw, fill=blue!20, align=center,
    text badly centered, node distance=6em, inner sep=0pt]
\tikzstyle{block} = [rectangle, draw, 
    text centered, align=center, node distance=4em]
\tikzstyle{iogram} = [trapezium, draw, fill=black!10, 
    trapezium left angle=70, trapezium right angle=-70,
    text centered, align=center, node distance=5em]
\tikzset{
  onslide/.code args={<#1>#2}{\only<#1>{\pgfkeysalso{#2}}}, 
}


\DeclareRobustCommand\suppOrApp{%
  \ifthenelse{\boolean{thesisStyle}}%
    {}%
    {Supplementary}%
}
\DeclareRobustCommand\Fig{%
  \ifthenelse{\boolean{thesisStyle}}%
    {Figure}%
    {Fig.}%
}

\DeclareRobustCommand\Figs{%
  \ifthenelse{\boolean{thesisStyle}}%
    {Figures\:}%
    {Figs.\:}%
}


\ifthenelse{\boolean{thesisStyle}}{%
  \floatstyle{plaintop}
  \restylefloat{table}
}
{}


\newcommand\D{\mathrm{d}}

\newcommand{\ANDw}{\land}
\newcommand{\ORw}{\lor}

\newcommand{\introSameGeneCredit}{%
\ifthenelse{\boolean{thesisStyle}}{%
  \footnote{This chapter is taken from material in \citealt{Shestov2013a}.
Brandon Barker is the primary author of all material found herein.}%
}%
{}%
}


\newcommand{\epiSameGeneCredit}{%
\ifthenelse{\boolean{thesisStyle}}{%
  \footnote{This chapter is published as \citet{Xu2012}.
Brandon Barker and Lin Xu contributed equally to this work.
It is additionally available in \citet[chapter 4]{Xu2012a}.}%
}%
{}%
}

\newcommand{\epiSameGeneAbstract}{%
Epistasis refers to the phenomenon in which phenotypic consequences
caused by mutation of one gene depend on one or more mutations at
another gene. Epistasis is critical for understanding many genetic and
evolutionary processes, including pathway organization, evolution of
sexual reproduction, mutational load, ploidy, genomic complexity,
speciation, and the origin of life. Nevertheless, current
understandings for the genome-wide distribution of epistasis are
mostly inferred from interactions among one mutant type per gene,
whereas how epistatic interaction partners change dynamically for
different mutant alleles of the same gene is largely unknown. Here we
address this issue by combining predictions from flux balance analysis
and data from a recently published high-throughput experiment. Our
results show that different alleles can epistatically interact with
very different gene sets. Furthermore, between two random mutant
alleles of the same gene, the chance for the allele with more severe
mutational consequence to develop a higher percentage of negative
epistasis than the other allele is 50-70\% in eukaryotic organisms,
but only 20-30\% in bacteria and archaea. We developed a population
genetics model that predicts that the observed distribution for the
sign of epistasis can speed up the process of purging deleterious
mutations in eukaryotic organisms. Our results indicate that epistasis
among genes can be dynamically rewired at the genome level, and call
on future efforts to revisit theories that can integrate epistatic
dynamics among genes in biological systems\epiSameGeneCredit.
}


\newcommand{\epistasisEnviroAbstract}{%
Epistasis describes the phenomenon that mutations at different loci do
not have independent effects with regard to certain
phenotypes. Understanding the global epistatic landscape is vital for
many genetic and evolutionary theories. Current knowledge for
epistatic dynamics under multiple conditions is limited by the
technological difficulties in experimentally screening epistatic
relations among genes. We explored this issue by applying flux balance
analysis to simulate epistatic landscapes under various
environmental perturbations. Specifically, we looked at gene-gene
epistatic interactions, where the mutations were assumed to occur in
different genes. We predicted that epistasis tends to become more
positive from glucose-abundant to nutrient-limiting conditions,
indicating that selection might be less effective in removing
deleterious mutations in the latter. We also observed a stable core of
epistatic interactions in all tested conditions, as well as many
epistatic interactions unique to each condition. Interestingly, genes
in the stable epistatic interaction network are directly linked to
most other genes whereas genes with condition-specific epistasis form
a scale-free network. Furthermore, genes with stable epistasis tend to
have similar evolutionary rates, whereas this co-evolving relationship
does not hold for genes with condition-specific epistasis. Our
findings provide a novel genome-wide picture about epistatic dynamics
under environmental perturbations.
}

\newcommand{\epistasisEnviroAuthorSummary}{%
Epistasis, often referred to as genetic interactions, occur when
mutational effects of genes depend on each other. Aside from often
times complicating the way in which the phenotype of an organism
relates to its genotype, epistatic interactions (or epistases) are
essential to several important theories in biology, especially in
evolution. Due to the difficulty in experimentally assessing epistasis
across an entire genome, we employed mathematical modeling of the
metabolic network of baker's yeast to comprehensively simulate genetic
interactions for virtually all known metabolic genes in the
organism. We performed comprehensive simulations in 17 different
environments, which differ by their nutrients. We characterized a
trend that occurs in genetic interactions when yeast is transferred
from a glucose-abundant environment to other environments. We also
found that both the set of genetic interactions present in all
conditions and the set of interactions present in a single environment
are fairly large sets with highly different connectivity. Furthermore,
the set present in all conditions tends to consist of gene pairs with
similar evolutionary rates.
}


\newcommand{\falconAbstractMotivation}{%
A major theme in constraint-based modeling is unifying 
experimental data, such as biochemical information about the reactions
that can occur in a system or the composition and localization of enzyme
complexes, with high-throughput data including expression data,
metabolomics, or DNA sequencing. The desired result is to increase
 predictive capability and improve our understanding of metabolism.
 The approach typically employed when only gene (or protein) intensities
are available is the creation of tissue-specific models, which reduces
the available reactions in an organism model, and does not provide an
objective function for the estimation of fluxes.
}

\newcommand{\falconAbstractResults}{%
We develop a method, flux assignment with LAD (least absolute
deviation) convex objectives and normalization (FALCON),
 that employs metabolic network reconstructions along with expression
data to estimate fluxes. In order to use such a method, accurate
measures of enzyme complex abundance are needed, so we first
present an algorithm that addresses quantification of complex
abundance. Our extensions to prior techniques include the
capability to work with large models and significantly improved
run-time performance even for smaller models, an improved analysis of
enzyme complex formation, the ability to handle large enzyme
complex rules that may incorporate multiple isoforms, and either
maintained or significantly improved correlation with experimentally
measured fluxes.
}

\newcommand{\falconAbstractAvail}{%
FALCON has been implemented in MATLAB and ATS, and can be downloaded
from: \url{https://github.com/bbarker/FALCON}. ATS is not required to
compile the software, as intermediate C source code is available. 
FALCON requires use of the COBRA Toolbox, also implemented in MATLAB.
}


\newcommand{\epiBeneMutAbstract}{%
Existing literature has only dealt with the simulation of strictly
deleterious mutants rather than beneficial mutants in the
constraint-based modeling literature, which is chiefly due to existing
studies optimizing the fitness function, which leaves no room for
improvement. In this study, we develop a constraint-based approach
that can simulate beneficial, neutral, and deleterious mutations.  We
show that this simulation technique can be useful for understanding
adaptive trajectories, and we develop a software library for the
analysis of evolutionary paths.  Our mechanistic model can reproduce
the distribution of
epistases between beneficial mutations that was observed in a data set
and a population genetic model fit to the same data set, showing that
our model behaves appropriately in this conext and may be a useful
tool for further evolutionary analyses. Finally, in experimental data
sets and in our simulations, slightly beneficial mutations are much
more likely to have positive (synergistic) epistasis with other
beneficial mutations, making their likelihood of becoming fixed higher
than would be expected without considering epistatic effects.
}

\captionsetup{labelfont=bf}




%% file: common/falcon.tex
\section{Introduction}

%
%
FBA (flux balance analysis) is the oldest, simplest, and perhaps
most widely used linear constraint-based metabolic modeling approach
\citep{Shestov2013a,Lewis2012}. FBA has become extremely popular, in
part, due to its simplicity in calculating reasonably accurate
microbial fluxes or growth rates
(e.g.\ \citealt{Schuetz2012,Fong2004_sb2013}); for
many microbes, a simple synthetic environment where all chemical
species are known suffices to allow proliferation, giving fairly
complete constraints on model inputs. Additionally, it has been found
that their biological objectives can be largely expressed as linear
objectives of fluxes, such as maximization of biomass \citep{Schuetz2012}. 
Neither of these assumptions necessarily hold for mammalian cells growing \textit{in
  vitro} or \textit{in vivo}, and in particular the environment is far
more complex for mammalian cell cultures, which have to undergo
gradual metabolic adaptation via titration to grow on synthetic media
\citep{Pirkmajer2011}. Recently, there have been many efforts to
incorporate both absolute and differential expression data into
metabolic models \citep{Blazier2012}. The minimization of metabolic
adjustment (MoMA; \citealt{Segre2002}) algorithm is the simplest
metabolic flux fitting algorithm, and it can be extended in order to
allow the use of absolute expression data for the estimation of flux
\citep{Lee2012}, which is the approach taken in this study. A different
approach for using expression in COBRA, also very simple, is E-flux,
which simply uses some function of expression (chosen at the
researcher's discretion; typically a constant multiplier of
expression) as flux constraints \citep{Colijn2009}. Despite this
surprising simplicity, the method has found many successful
applications, but the user-chosen parameter and use of expression as
hard constraints is, in our opinion, a detraction, and others have had
better results taking an approach similar to \citealt{Lee2012}
\citep{Bogart2015}.


The MoMA method, framed as a constrained least-squares optimization
problem, is typically employed to calculate the flux vector of an
\textit{in silico} organism after a mutation by minimizing the distance
between the wild-type flux and the mutant flux. The biological
intuition is that the organism has not had time to adapt to the
restricted metabolic capacity and will maintain a similar flux to the
wild-type (WT) except where the perturbations due to the mutation
dictate necessary alterations in fluxes \citep{Shlomi2005}. Suppose
$\mathbf{a}$ is the WT flux vector obtained by an optimization
procedure such as FBA, empirical measurements, or a combination of
these. For an undetermined flux vector $\mathbf{v}$ in a model with
$N$ reactions the MoMA objective can be expressed as

\begin{equation}
\centering
\textnormal{minimize}\ \sum\limits_{i=1}^N (v_i-a_i)^2 
\end{equation}

subject to the stoichiometric constraints $\mathbf{S v}\nolinebreak
=\nolinebreak \mathbf{0}$ where $\mathbf{v} = (v_1, \ldots,
v_N)^T$ and $\mathbf{S}$ is the stoichiometric matrix (rows correspond
to metabolites, columns to reactions, and entries to stoichiometric
coefficients). Constant bounds on fluxes are often present, such as
substrate uptake limits, or experimental $V_{\max}$ estimates, so we
write these as the constraints $\mathbf{v}_{lb}\nolinebreak
\preceq\nolinebreak \mathbf{v}\nolinebreak \preceq \mathbf{v}_{ub}$.
The objective may be equivalently expressed in the canonical
quadratic programming (QP) vector form as
$\textnormal{min.\ }\ \frac{1}{2}\mathbf{v}^T \mathbf{v}\nolinebreak
-\nolinebreak \mathbf{a}^T \mathbf{v}$. This assumes that each $a_i$
is measured, but it is also possible and sometimes even more useful to
employ this objective when only a subset of the $a_i$ are measured (if
$a_i$ is not measured for some $i$, then we omit $(v_i-a_i)^2$ from
the objective). In metabolomics, for instance, it is always the case
in experiments with labeled isotope tracers that only a relatively
small subset of all fluxes are able to be estimated with metabolic
flux analysis (MFA; \citealt{Shestov2013a}). Combining MoMA with MFA
provides a technique to potentially estimate other fluxes in the
network.

A variant of MoMA exists that minimizes the absolute value of the
difference between $a_i$ and $v_i$ for all known $a_i$. To our
knowledge, the following linear program is the simplest version of
linear MoMA, which assumes the existence of a constant flux vector
$\mathbf{a}$:

\begin{equation}
\centering
\begin{tabular}{rl}
minimize & $\sum\limits_{i=1}^N d_i$  \\
subject to & $\mathbf{S v} = \mathbf{0}$ \\
 & $\mathbf{v}_{lb} \preceq \mathbf{v} \preceq \mathbf{v}_{ub}$ \\
$\forall i:$ & $-d_i \le v_i-a_i \le d_i$ \\
 & $d_i \ge 0$
\end{tabular}
\end{equation}

The $d_i$ are just the distances from \textit{a priori} fluxes to
their corresponding fitted fluxes. Linear MoMA has the advantage that
it is not biased towards penalizing large magnitude fluxes or
under-penalizing fluxes that are less than one
\citep{Boyd2004,Shlomi2005}. Additionally, linear programs are often
amenable to more alterations that maintain convexity than a quadratic
program and tend to have fewer variables take on small values, and it
is much easier to interpret the importance of a zero than 
a small value \citep{Boyd2004}.

We wish to apply MoMA to expression data rather than flux data, but there
are two primary problems that must be tackled. First, we must quantify
enzyme complex abundance as accurately as possible given the gene expression
data. Although there is not a one-to-one correspondence between
reactions and enzyme complexes, the correspondence is much closer
than that between individual genes and metabolic reactions. In the
first part of this work, we employ an algorithm that can account for
enzyme complex formation and thus quantify enzyme complex
abundance. Second, we must fit real-valued variables (fluxes) to 
non-negative data (expression), which is challenging to do
efficiently. To accomplish this, we build on the original MoMA
objective, which must be altered in several ways (also discussed in
\citealt{Lee2012}, which lays the groundwork for the current
method). We develop automatic scaling of expression values so that
they are comparable to flux units obtained in the optimization
routine. This can be an advantage over the prior method as it no longer requires
the manual choice of a flux and complex abundance pair with a ratio that
is assumed to be representative of every such pair in the system.
Related to this, we also implement the sharing of enzyme complex
abundance between the reactions that the complex catalyzes, rather
than assuming there is no competition between reactions catalyzed by
the same complex. Reaction direction assignment enables comparison of
fluxes and expression by changing fluxes to non-negative values. We
show that batch assignment, rather than serial assignment
\citep{Lee2012} of reaction direction can greatly improve time
efficiency while maintaining or slightly improving correlations
with experimental fluxes. In addition to several of the methods 
described so far, we also included in our comparison two methods for
\emph{tissue-specific} modeling. In GIMME, the authors remove
reactions whose associated gene expression is below some threshold,
then add reactions that preclude some user-defined \emph{required
metabolic functionalities} in an FBA objective back into the model,
and finally use FBA again to obtain fluxes \citep{Becker2008}. The
other tissue-specific method we compared with is iMAT, which employs a
mixed integer linear programming (MILP) problem to maximize the number
of reactions whose activity corresponds to their expression state
(again using thresholds, but this time, there are low, medium, and
highly expressed genes, and only the lowly and highly expressed genes
are included in the objective) all while subject to typical
constraints found in FBA \citep{Shlomi2008}.
 
Finally, we employ several sensitivity analyses and
performance benchmarks so that users of the FALCON method and related
methods may have a better understanding of what to expect in practice.
\section{Methods}

\begin{figure}
\centering
\begin{tikzpicture}
    \node [block] (start) {start}; 
    \node [iogram, below of=start, left of=start] (exp) {Genes:\\ expression 
      ($\mu$,~$\sigma$)}; 
    \node [iogram, below of=start, right of=start] (rules) {Reactions:
      GPR Rule}; 
    \node [block, below of=rules] (parse) {Parse Rule}; 
    \node [block, below of=parse, left of=parse, xshift=-0.5cm]
      (mindisj) {Find minimum disjunction};
    \node [iogram, below of=mindisj] (expstd)
          {Reactions (enzyme~complexes):\\ abundance ($\mu$,~$\sigma$)};
    \node [iogram, right of=falcon, below of=expstd] (smat) 
      {$\mathbf{S}$ matrix};
    \node [iogram, left of=falcon, below of=expstd] (vbnd) 
      {Reactions:\\ flux bounds ($\mathbf{v}_{lb}$, $\mathbf{v}_{ub}$)};
    \node [block, below of=vbnd, left of=smat, right of=vbnd, 
      below of=expstd, yshift=-0.5cm] (falcon) 
      {Flux fitting (FALCON)};
    \node [iogram, below of=falcon] (fluxout) 
      {Reactions:\\ fluxes ($\mathbf{v}:$ $\mu$,~$\sigma$)};
    \path [line] (start) -- (exp);
    \path [line] (start) -- (rules);
    \path [line] (rules) -- (parse);
    \path [line] (exp.south) -- (mindisj);
    \path [line] (parse) -- (mindisj);
    \path [line] (mindisj) -- (expstd);
    \path [line, transform canvas={xshift=0.25cm}] (expstd) -- (falcon);
    \path [line] (rules.east) |- (falcon.east);
    \path [line] (smat) -- (falcon);
    \path [line] (vbnd) -- (falcon);
    \path [line] (falcon) -- (fluxout);

\end{tikzpicture}
\caption{Flowchart illustrating the two algorithms used in this
paper. The process of estimating enzyme complex abundance is displayed
in detail, whereas the flux-fitting algorithm (FALCON) is illustrated
as a single step for simplicity. First, for each gene in the model
with available expression data, the mean and (if available) standard
deviation or some other measure of uncertainty are read in. Gene rules
(also called GPR rules) are also read in for each enzymatic
reaction. The reaction rules are parsed and the minimum disjunction
algorithm is applied, making use
of the gene's mean expression. Next, the estimated and unitless enzyme
complex abundance and variance are output for each enzymatic
reaction. Finally, flux fitting with FALCON
(Algorithm~\ref{alg:FALCON}) can be applied, and requires the model's
stoichiometry and flux bounds. The final output has the option of
being a deterministically estimated flux, or a mean and standard
deviation of fluxes if alternative optima are explored.}
\label{ECCN_flowchart}
\end{figure}
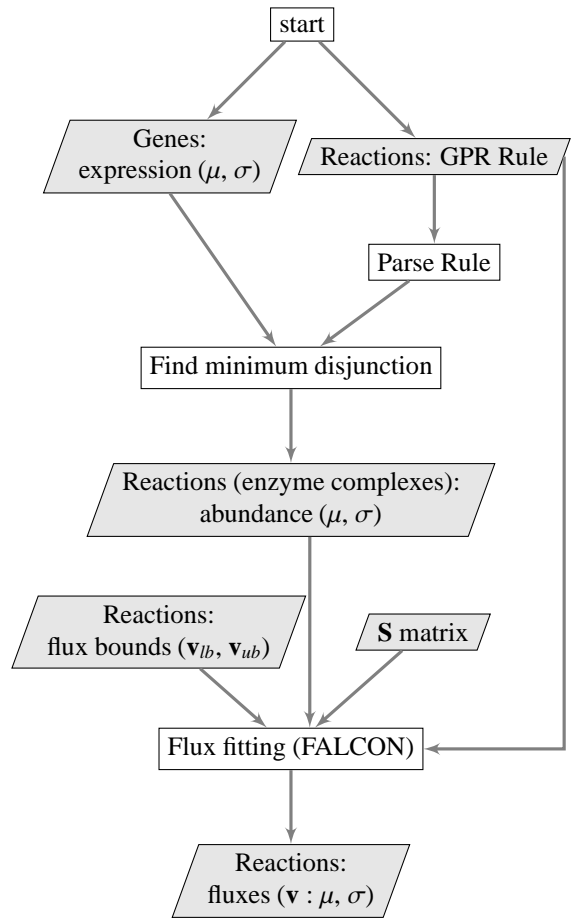

Most genome-scale models have attached Boolean (\textit{sans}
negation) gene rules to aid in determining whether or not a gene
deletion will
completely disable a reaction. These are typically called GPR
(gene-protein-reaction) rules and are a requirement for FALCON; their
validity, like the stoichiometric matrix, is
important for generating accurate predictions. Also important are the
assumptions and limitations for the process of mapping expression data
to complexes so that a scaled enzyme complex abundance (hereafter
referred to as complex abundance) can be estimated. We address these
in the next section and have attached a flow chart to illustrate the
overall process of mapping expression of individual genes to enzyme
complexes within the greater context of flux estimation 
(\Fig~\ref{ECCN_flowchart}). We employ an algorithm for
this step---finding the minimum disjunction---for estimating complex
abundance as efficiently and as accurately as possible given the
assumptions (\suppOrApp Section~\ref{sec:complexation}).

Consideration of constraint availability, such as assumed reaction
directions and nutrient availability, is crucial in this type of
analysis. In order to work with two sets of constraints with
significantly different sizes in yeast, we
wrote the MATLAB function \texttt{removeEnzymeIrrevs} to find all
enzymatic reactions in a model that are annotated as reversible but
are constrained to operate in one direction only. This is not
something a researcher would normally want to do since constraints
should generally act to improve modeling predictions, but we are
interested to see how their removal influences model predictions and
solution robustness.

The function
\texttt{useYN5irrevs} copies the irreversible annotations found in
Yeast~5.21 \citep{Lee2012} to a newer yeast model, but could in
principle be used for any two models; by default, this script is coded
to first call \texttt{removeEnzymeIrrevs} on both models before
copying irreversible annotations. Application of these scripts removes
853 constraints in Yeast~5.21 and 1,723 constraints in Yeast~7.
Despite the significant relaxation in constraints, since nutrient
uptake constraints are unaffected, FBA only predicts a 1.28\% increase
in growth rate in the minimally constrained Yeast~7 model. However, in
FALCON, we are no longer optimizing a sink reaction like biomass, and
this relaxation in internal constraints proves to be more important.
Constraint sets for Human Recon~2 are described in \suppOrApp 
\Fig~\ref{fig:ExpSensRec2}.

\subsection{Estimating enzyme complex abundance}

Given the diversity and availability of genome-scale expression datasets,
either as microarray or more recently RNA-Seq, it could be useful to
gauge the number of enzyme complexes present in a cell. A recent
study found that only 11\% of annotated \textit{Drosophila}
protein complexes have subunits that are co-expressed
\citep{Juschke2013}, so it cannot be assumed that any given protein
subunit level represents the actual complex abundance. We formalize a
model for enzyme complex formation based on GPR rules that are
frequently available in genome-scale annotations.

The original expression to complex abundance mapping procedure \citep{Lee2012}
performed a direct evaluation of GPR rule expression
values---replacing gene names with their expression values, ANDs with
minimums, and ORs with sums, without altering the Boolean expression
of the GPR rule in any way. Below we illustrate a 
problem that can occur with this mapping where some genes' expression
levels may be counted more than once. 

The $r_i$ are different reaction rules and the $e_i$ are the
corresponding estimated complex abundance levels. Lower case letters
are shorthand for the expression level of the corresponding gene ID in
uppercase; for example, $a = \E{\textnormal{A}}$, where
$\E{\textnormal{A}}$ is the expression of gene A.

\begin{equation}
\centering
{\setlength{\tabcolsep}{.16667em}
\begin{tabular}{cccccccc}
$r_1$ & := & [A and B] or [A and C] & $\rightarrow$ & $e_1$  &=& $\min(a,b$) + $\min(a,c$) \\ 
$r_2$ & := & [A and (B or C)]       & $\rightarrow$ & $e_2$  &=&  $\min(a, b + c$) 
\end{tabular} 
}
\end{equation}

Supposing A is the minimum, then if we just evaluate $r_1$ directly (a
rule in disjunctive normal form, or DNF), A will be counted twice.
Rules with sub-expressions in DNF are frequently encountered in practice,
but directly evaluating them can lead to erroneous quantification.

Another possibility is partitioning expression among multiple occurrences
of a gene in a rule. For instance, in $r_1$ above, we could evaluate
it as $e_1$ =
min($\frac{a}{2},b$) + min($\frac{a}{2},c$) to account for the
repeated use of $a$. However, other potential issues aside, we can see
that this can cause problems rather quickly. For instance, suppose $b
= a$ and $c = 0$; then min($a$,$b+c$) $=b=a$ appears to be correct,
not min($\frac{a}{2},b$) + min($\frac{a}{2},c$) = $\frac{a}{2} + 0$.
From this example, we can see that conversion to conjunctive normal
form (CNF; \citealt{Russell2009}), as in $r_2$ appears to be a
promising prerequisite for evaluation.

\subsection{The min-disjunction algorithm estimates enzyme complex abundance}

In \suppOrApp Section~\ref{sec:complexation}, we show that
converting a rule to CNF is a sound method
to aid in the estimation of enzyme complex abundance. The minimum disjunction
algorithm is essentially just the standard CNF conversion algorithm \citep{Russell2009},
with the implementation caveat that a gene that is in disjunction with itself should be
reduced to a literal. We've found that this makes the CNF conversion algorithm tractable
for all rules and prevents double counting of gene
expression. Conversion to CNF and selection of the minimum disjunction
also removes redundant genes from the complex (e.g. holoenzymes; see
\suppOrApp Assumption~\ref{asm:holo}). Biologically, selecting the
minimum disjunction effectively finds the \emph{rate-limiting}
component of enzyme-complex formation. After conversion to CNF, the
minimum disjunction algorithm substitutes gene-expression values as
described in \citealt{Lee2012} and evaluates the resulting arithmetic
expression. Another new feature of our approach is the handling of
missing gene data. If expression is not measured for a gene in a GPR
rule, the rule is modified so that the missing gene is no longer part
of the Boolean expression. For instance, if data is not measured for
gene B in [A and (B or C)] then the rule would become [A and C]. This
prevents penalization of the rule's expression value in the case that
the missing gene was part of a conjunction, and it also assumes there
was no additional expression from the missing gene if it is in a
disjunction.

Although conversion to CNF may be intractable for some expressions
\citep{Russell2009}, we tested our implementation of the algorithm on
three of the most well-curated models which contain some of the
most complex GPR rules available. These models are for \textit{E. coli}
\citep{Orth2011}, yeast \citep{Aung2013}, and human
\citep{Thiele2013}. In all cases, the rules were converted to CNF in
less than half a second, which is far less than the typical flux
fitting running time from Algorithm~\ref{alg:FALCON}.

Using the minimum disjunction method results in several
differences from direct substitution and evaluation in yeast GPR
rules. When data completely covers the genes in the model
(e.g. \citealt{Lee2012}), complex abundance tends to have few differences in yeast
regardless of the evaluation method (25 rules; 1.08\% of all rules for
Yeast~7). This number goes up significantly in Human Recon~2
\citep{Thiele2013} due to more complex GPR rules (935 rules; 22\% of
all rules). For the human model, we could not find any data set that
covered every gene, so instead random expression data roughly matching
a power law was used to generate this statistic. If we use proteomics
data for yeast and human models, the algorithmic variation in how missing gene
data is handled causes some additional increase in differences
\citep{Picotti2013,Gholami2013}.  For proteomics, in the Yeast~7 model
205 rules (8.87\% of all rules) differed, and in Human Recon~2, 1002
rules (23.57\% of all rules) differed. We can see that for yeast, the
changes in flux attributed to enzyme abundance evaluation can be
relatively small for data with 100\% gene coverage, but can be 
significant in Human (\suppOrApp \Fig~\ref{fig:EnzAbundEval}).

%
%

\section{The FALCON algorithm}
\label{sec:FALCON}

Prior work that served as an inspiration for this method used Flux
Variability Analysis (FVA) to determine reaction direction
\citep{Lee2012}. Briefly, this involves two FBA simulations per
reaction catalyzed by an enzyme, and as the algorithm is iterative,
this global procedure may be run several times before converging to a
flux vector. We removed FVA to mitigate some of the cost, and instead
assign flux direction in batch; while it is possible that the
objective value may decrease in subsequent iterations of the
algorithm, this is not an issue since the objective function is
changed at each iteration to include more irreversible
fluxes. The objective value of a function with more
fluxes should supersede the importance of one with fewer fluxes, as
the functions are not the same and thus not directly comparable, 
and we wish to include as many data points in the fitting as possible.

\begin{figure}
\centering
\begin{tabular}{c}
  \begin{subfigure}[b]{0.6\textwidth}
  \includegraphics[width=\textwidth, height=0.4\textheight]
  {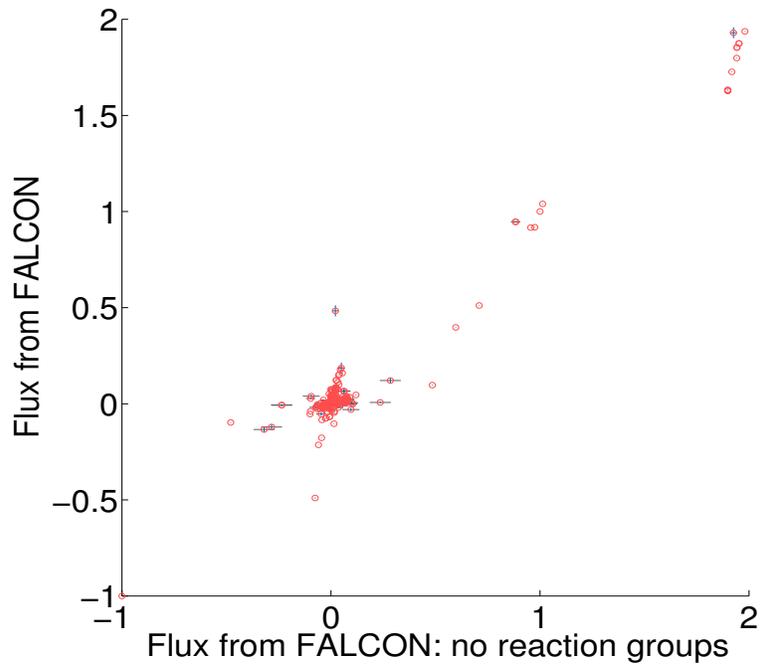}
  \caption{} \label{fig:FalconGrp:A}
  \end{subfigure}
\\
  \begin{subfigure}[b]{0.6\textwidth}
  \includegraphics[width=\textwidth, height=0.4\textheight]
  {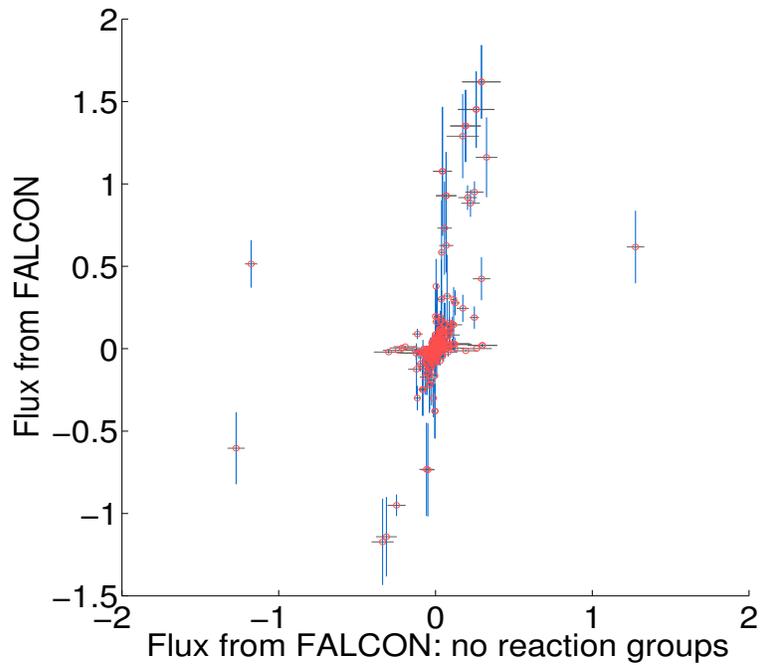}
  \caption{} \label{fig:FalconGrp:B}
  \end{subfigure} 
\\
\end{tabular}
\caption{
Comparison of setting FALCON to use no reaction group information (x-axis)
versus with group information (y-axis; default FALCON setting) for
both the highly constrained Yeast~7 model \textbf{(a)} and the
minimally constrained Yeast~7 model \textbf{(b)}.
Error bars with length equal to one standard deviation are
shown for both approaches as a result of alternative solutions in
FALCON.}
\label{fig:FalconGrp}
\end{figure}

One major advance in our method is the consideration of enzyme
complexes sharing multiple reactions, which we call reaction groups.
This is done by partitioning an enzyme complex's abundance across
its reactions by including all reactions associated to the complex
in the same constraint.
Both minimally and highly constrained models
(Section~\ref{sec:sensToExpNoise}) show some fluxes with significant
differences depending on the use of group information, particularly in
the minimally constrained model (\Fig~\ref{fig:FalconGrp}). We now
discuss the algorithm in detail, including several
other important features, including automatic scaling of
expression.

To make working with irreversible fluxes simpler, we convert the model
to an irreversible model, where each reversible flux $v_j$ in the
original model is split into a forward and a backward reaction that take
strictly positive values: $v_{j,f}$ and $v_{j,b}$. We also account for
enzyme complexes catalyzing multiple reactions by including all
reactions with identical GPR rules in the same residual
constraint; indexed sets of reactions are denoted $R_i$ and their
corresponding estimated enzyme abundance is $e_i$. \Fig~\ref{fig:FalconGrp} 
shows the difference in Algorithm~\ref{alg:FALCON} when we do not use
reaction group information.  The standard deviation of enzyme
abundance, $\sigma_i$, is an optional weighting of uncertainty in
biological or technical replicates.

We employ a normalization variable $n$ in the problem's objective and
flux-fitting constraints to find the most agreeable scaling of
expression data. The linear fractional program shown below can be
converted to a linear program by the Charnes-Cooper transformation
\citep{Boyd2004}. To avoid the need for fixing any specific flux, which
may introduce bias, we introduce the bound $\sum_{j \,\mid\, \exists i
\textnormal{{ }s.t.{ }} j\in R_i} \left|v_j\right| \geq
V_{lb}^{\Sigma}$. This guarantees that
the optimization problem will yield a non-zero flux vector. As an
example of how this can be beneficial, this means we do not need to
measure any fluxes or assume a flux is fixed to achieve good results;
though this does not downplay the value of obtaining
experimentally-based constraints on flux when available 
(\suppOrApp \Fig~\ref{fig:FluxBars}).

The actual value of $V_{lb}^{\Sigma}$ is not very important due to the
scaling introduced by $n$, and we include a conservatively small value
that should work with any reasonable model. However, for numeric
reasons, it may be best if a user chooses to specify a value
appropriate for the model. Similarly, if any fluxes are known or
assumed to be non-zero, this constraint becomes unnecessary. To keep
track of how many reactions are irreversible in the current and prior
iteration, we use the variables $rxns_{irrev}$ and
$rxns_{irrev,prior}$. The algorithm terminates when no reactions are
constrained to be exclusively forward or backward after an iteration.

\begin{algorithm}
\caption{FALCON}
\label{alg:FALCON}
\begin{algorithmic}
\INPUT $\{R_i : i\ $ an index for a unique enzyme complex where\\ 
        $\ R_i = \{ j: \textnormal{{}complex}\ i \textnormal{{ }catalyzes reaction{ }} j\}\}
       $
\INPUT enzyme abundances (mean: $e_i$, standard deviation: $\sigma_i$)
\INPUT model ($\mathbf{S}$ matrix, $\mathbf{v}_{lb}$, $\mathbf{v}_{ub}$)
\State $u_{\min} \leftarrow \min_j\ \{V_{j,\max} : V_{j,\max} > 0\}$
       where $V_{j,\max} = \max\left(\left|v_{lb,j}\right|, \left|v_{ub,j}\right|\right)$.
\State $V_{lb}^{\Sigma} \leftarrow u_{\min} \left|\{j : \exists i
       \textnormal{{ }s.t.{ }} j\in R_i\}\right|$
\State $rxns_{irrev} \leftarrow $ number of reactions ($j$) such that either 
$v_{ub,j} > 0$ or $v_{lb,j} < 0$, but not both. 
\ForAll {i}
  \State {Scale data to be of similar size for numeric stability:}
  \State $e_i \leftarrow \frac{e_i V_{lb}^{\Sigma}}
    {\sum\limits_{j} e_i}$ 
  \State $\sigma_i \leftarrow \frac{\sigma_i V_{lb}^{\Sigma}}
    {\sum\limits_{j} e_i}$
\EndFor
\While{$rxns_{irrev} > rxns_{irrev,prior}$}
  \State {$rxns_{irrev,prior} \leftarrow rxns_{irrev}$}
  \State Call LP Solver (updates $\mathbf{v}$):
  \State \hspace{4.8mm} \parbox[t]{\dimexpr\linewidth-\algorithmicindent}{
    $\textnormal{minimize}\ \sum\limits_i \frac{d_i}{n \sigma_i}$\\
    s.t.\\
    $\sum_{j \,\mid\,\exists i \textnormal{{ }s.t.{ }} j\in R_i} \left|v_j\right| 
      \geq V_{lb}^{\Sigma}$\\ 
    $\forall i: -d_i \leq \sum\nolimits_{j \in R_i} (v_{j,f} +
    v_{j,b}) - n e_i \leq d_i$ where $v_j = v_{j,f} - v_{j,b}$\\
    $d_i, v_{j,f}, v_{j,b} \geq 0$\\ 
    $n > 0$
    \strut}
  \ForAll {$\left\{j \mid v_{j,f} + v_{j,b} > 0, v_{j,f} \neq v_{j,b} \right\}$}
  \State {Constrain the smaller of $v_{j,f}$ and $v_{j,b}$ to be $0$.}  
  \State {$rxns_{irrev} \leftarrow rxns_{irrev} + 1$}
  \EndFor
\EndWhile
\OUTPUT $\mathbf{v}$
\end{algorithmic}
\end{algorithm}

Algorithm~\ref{alg:FALCON} and the method in \citealt{Lee2012} are
both non-deterministic. In the first case, Algorithm~\ref{alg:FALCON}
solves an LP during each iteration, and subsequent iterations depend
on the LP solution, so that alternative optima may affect the outcome.
In the latter case, alternative optima of individual LPs is not an
issue, but the order in which reactions assigned to be irreversible can
lead to alternative solutions. However, we
found that the variation due to this stochasticity is typically 
relatively minor, particularly in cases where the model is more 
heavily constrained (\suppOrApp Figs. \ref{fig:EnzAbundEval} and 
\ref{fig:FluxBars}).






\section{Results and Discussion}

\subsection{Performance benchmarks}
Using the same yeast exometabolic and expression data employed for
benchmarking in the antecedent study \citep{Lee2012} that included an
updated version of the Yeast~5 model \citep{Heavner2012} and the latest
yeast model \citep{Aung2013}, we find that
our algorithm has significant improvements in time efficiency while
maintaining correlation with experimental fluxes, and is much faster 
than any similarly performing method (Table~\ref{tab:FalcPerf}; 
\suppOrApp \Fig~\ref{fig:FluxBars}). 
Timing for the human model also improved in FALCON; in a model with
medium constraints and exometabolic directionality constraints, FALCON
completed on average in 3.6~m and the method from \citealt{Lee2012} in
1.04~h. Furthermore, when we remove many bounds constraining the
direction of enzymatic reactions that aren't explicitly annotated as
being irreversible in prior work \citep{Lee2012}, we find that our
formulation of the approach seems to be more robust than other
methods.

\begin{table*}
\begin{center}
\resizebox{\textwidth}{!}{%
\begin{tabular}{lrrrrrrrrr}
\textbf{(a)} \hspace{1.2cm} & Max. $\mu$ & Model & Experimental & 
  Standard FBA & Fitted FBA & GIMME & iMAT & Lee et al. & FALCON\\
 & 75 \%& Yeast~5 MC & 1 & 0.66 & 0.66 & NaN  & 0.57 & 0.64 & 1 \\
 & 75 \%& Yeast~7 MC & 1 & 0.66 & 0.66 & 0.68 & 0.66 & 0.66 & 0.98\\
 & 75 \%& Yeast~5 HC & 1 & 0.73 & 0.78 & 0.75 & 0.66 & 0.98 & 0.99\\
Pearson's r  
 & 75 \%& Yeast~7 HC & 1 & 0.70 & 0.70 & 0.80 & 0.66 & 0.98 & 0.99\\
 & 85 \%& Yeast~7 MC & 1 & 0.62 & 0.62 & 0.65 & 0.62 & 0.62 & 0.97\\
 & 85 \%& Yeast~5 HC & 1 & 0.88 & 0.89 & 0.9  & 0.81 & 0.99 & 0.99\\
 & 85 \%& Yeast~7 HC & 1 & 0.67 & 0.67 & 0.87 & 0.62 & 0.98 & 0.98\\
\end{tabular}}
\resizebox{\textwidth}{!}{%
\begin{tabular}{lrrrrrrrrr}
\textbf{(b)} \hspace{1.2cm} & Max. $\mu$ & Model & Experimental & 
  Standard FBA & Fitted FBA & GIMME & iMAT & Lee et al. & FALCON\\
 & 75 \%& Yeast~5 MC & 0 & 0.9  & 470   & 0.81  & 50     & 110 & 1.8 \\
 & 75 \%& Yeast~7 MC & 0 & 1.9  & 3,100 & 2.1   & 12,000 & 600 & 5.6 \\
 & 75 \%& Yeast~5 HC & 0 & 0.12 & 110   & 0.18  & 1.4    & 15  & 0.27\\
Time (s) 
 & 75 \%& Yeast~7 HC & 0 & 0.72 & 940   & 1.7   & 240    & 670 & 5.5 \\
 & 85 \%& Yeast~7 MC & 0 & 2.3  & 3,100 & 3.8   & 14,000 & 610 & 4.6 \\
 & 85 \%& Yeast~5 HC & 0 & 0.12 & 110   & 0.18  & 2.5    & 15  & 0.22\\
 & 85 \%& Yeast~7 HC & 0 & 0.70 & 110   & 2.5   & 100    & 530 & 5.9\\
\end{tabular}}
\end{center}
\caption{Performance of FALCON and other CBM methods for predicting
yeast exometabolic fluxes in two growth conditions with highly (HC)
and minimally (MC) constrained models \textbf{(a)} and associated
timing analysis \textbf{(b)}. For Lee et al. and FALCON methods, the
mean time for a single run of the method is listed; all other methods
did not have any stochasticity employed. Values are shown in two
significant figures. Method descriptions can be found in
\protect\citealt{Lee2012}.}
\label{tab:FalcPerf}
\end{table*}

\begin{figure}
\centering
\begin{tabular}{c}
  \begin{subfigure}[b]{\textwidth}
  \includegraphics[width=\textwidth]{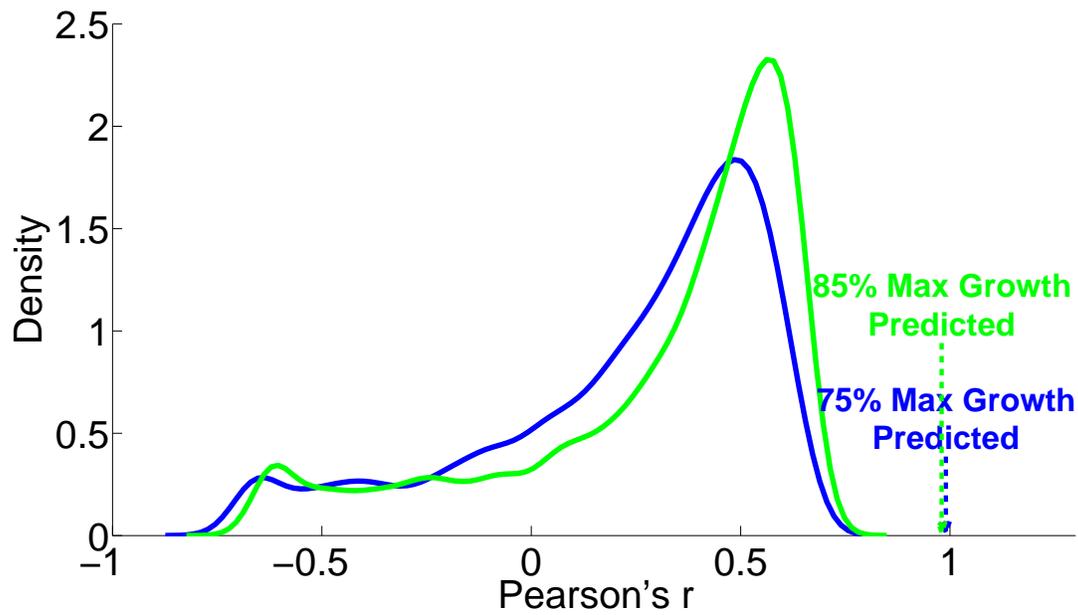}
  \caption{highly constrained} \label{fig:YpermCorr:A}
  \end{subfigure}
\\
  \begin{subfigure}[b]{\textwidth}
  \includegraphics[width=\textwidth]{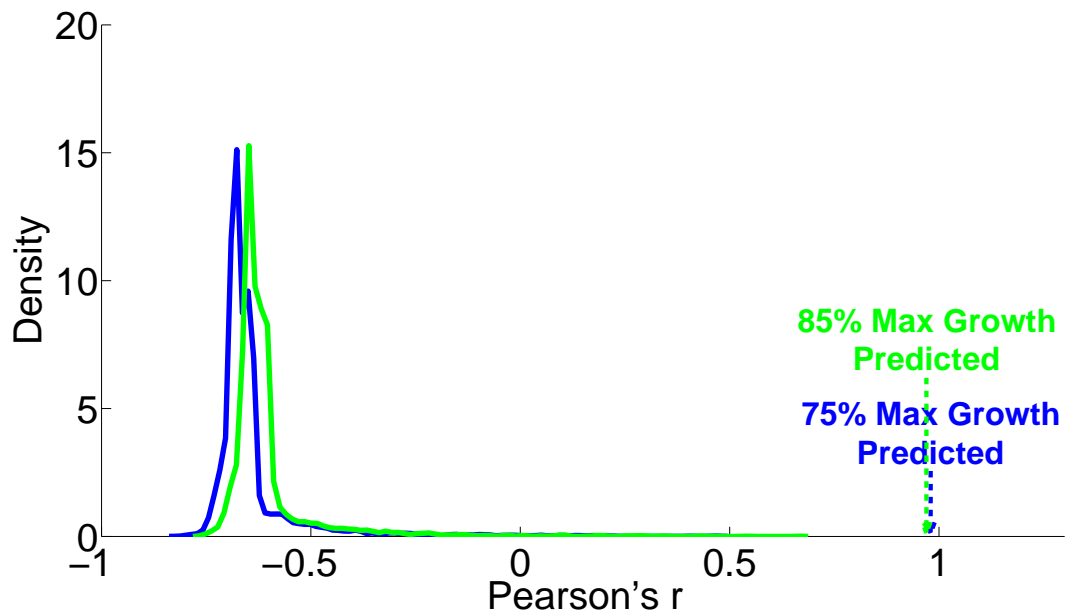}
  \caption{minimally constrained} \label{fig:YpermCorr:B}
  \end{subfigure} 
\\
\end{tabular}
\caption{Kernel-smoothed PDFs of correlation between experimental
fluxes and fluxes estimated from FALCON when all gene expression data
points are permuted. Arrows mark the correlation when FALCON is run on
the unpermuted expression data.  Random correlations tend to be much
more positive in the highly constrained model \textbf{(a)} than in the
minimally constrained model \textbf{(b)}. 5,000 permutation replicates
were performed in all cases.}
\label{fig:YpermCorr}
\end{figure}

We see that the predictive ability of the algorithm does not
appear to be an artifact; when FALCON is run on permuted expression data,
it doesn't do as well as the actual expression vector (\Fig~\ref{fig:YpermCorr}).
The full-sized flux vectors estimated from permuted expression as a
whole also does not correlate well with the flux vector estimated from
the actual expression data, but we notice that the difference is
visibly larger in the minimally constrained model compared to the
highly constrained model (\suppOrApp \Fig~\ref{fig:YpermCorrSup}). Rigidity in the
highly constrained model appears to keep most permutations from
achieving an extremely low Pearson correlation, likely due to forcing
fluxes through the same major pathways, but a rank-based correlation
still shows strong differences.
 
\subsection{Sensitivity to expression noise}
\label{sec:sensToExpNoise}
To understand the sensitivity of flux to expression, we multiply noise
from multivariate log-normal distributions with the expression vector
and see the effect on the estimated fluxes. For instance, correlation
between two types of proteomics data yields a Pearson's $r = 0.7$
\citep{Gholami2013}, corresponding to an expected $\sigma \approx 1.4$
and expected $r \approx 0.4$ for flux in our most highly constrained
human model (\suppOrApp \Fig~\ref{fig:ExpSensRec2}). We find that
enzymatic reaction directionality constraints influence the
sensitivity of the model to expression perturbation
(\Fig~\ref{fig:ExpSens}). It is important to note that mere presence
of the constraints does not help us determine the correct experimental
fluxes when other classes of methods (e.g.\ FBA;
Table~\ref{tab:FalcPerf}) are used. Additionally, it is possible to
obtain good predictions even without a heavily constrained model
(Table~\ref{tab:FalcPerf}).

With Human Recon~2, additional constraint sets supply some benefit,
but even the most extreme constraint set does not compare to what is
available in Yeast~7, which is also inherently constrained by the fact
that yeast models will be smaller than comparable human models
(\suppOrApp \Fig~\ref{fig:ExpSensRec2}). For mammalian models, more
sophisticated means of constraint, such as enzyme crowding constraints
\citep{Shlomi2011}, or using FALCON in conjunction with tissue
specific modeling tools, may prove highly beneficial.

\newlength{\expSensLen}
\setlength{\expSensLen}{0.7\textwidth}
\begin{figure}[H]
\centering
\begin{tabular}{c}
  \begin{subfigure}[b]{\expSensLen}
  \includegraphics[width=\textwidth]{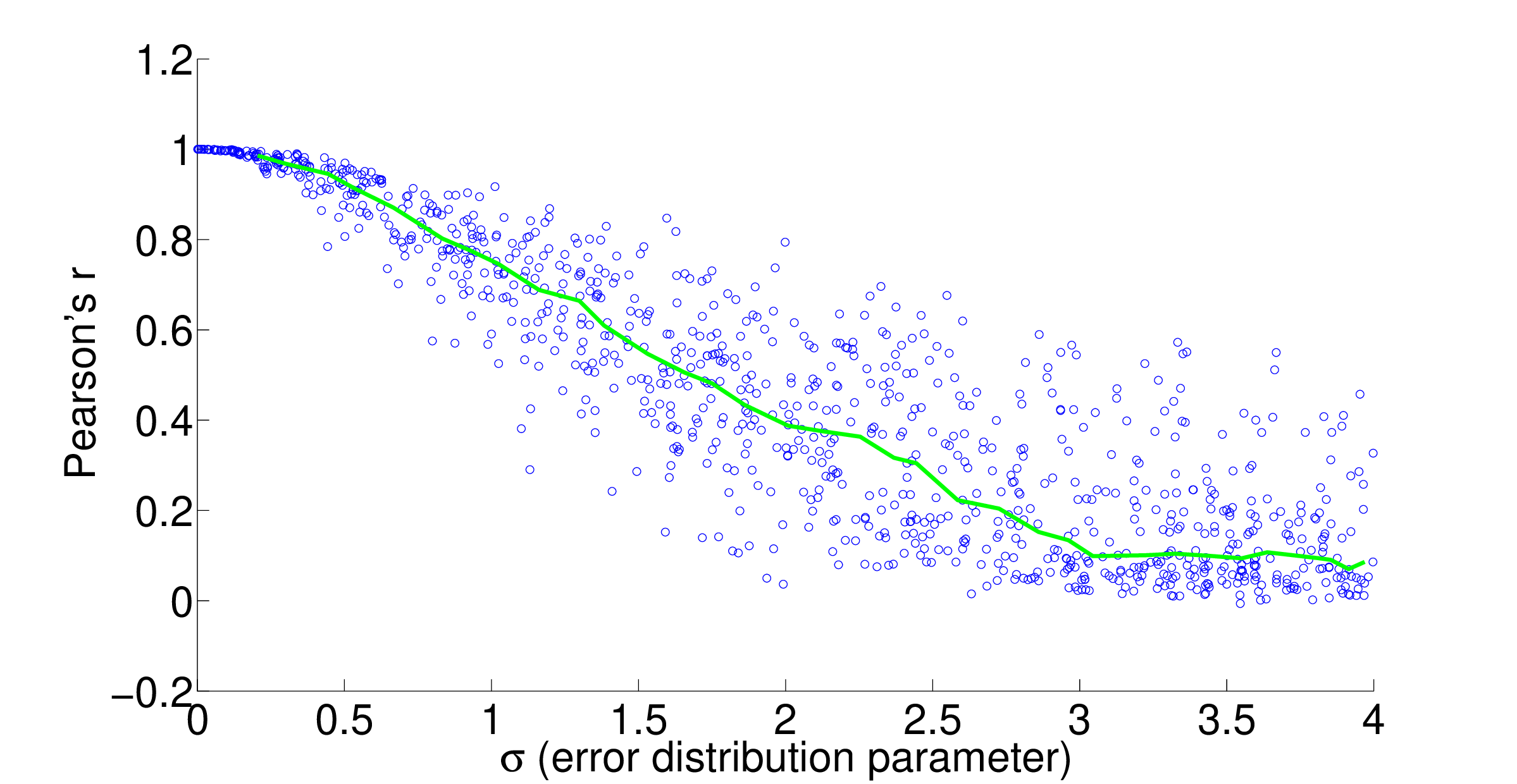}
  \caption{enzyme complex abundance} \label{fig:ExpSens:A}
  \end{subfigure}
\\
  \begin{subfigure}[b]{\expSensLen}
  \includegraphics[width=\textwidth]{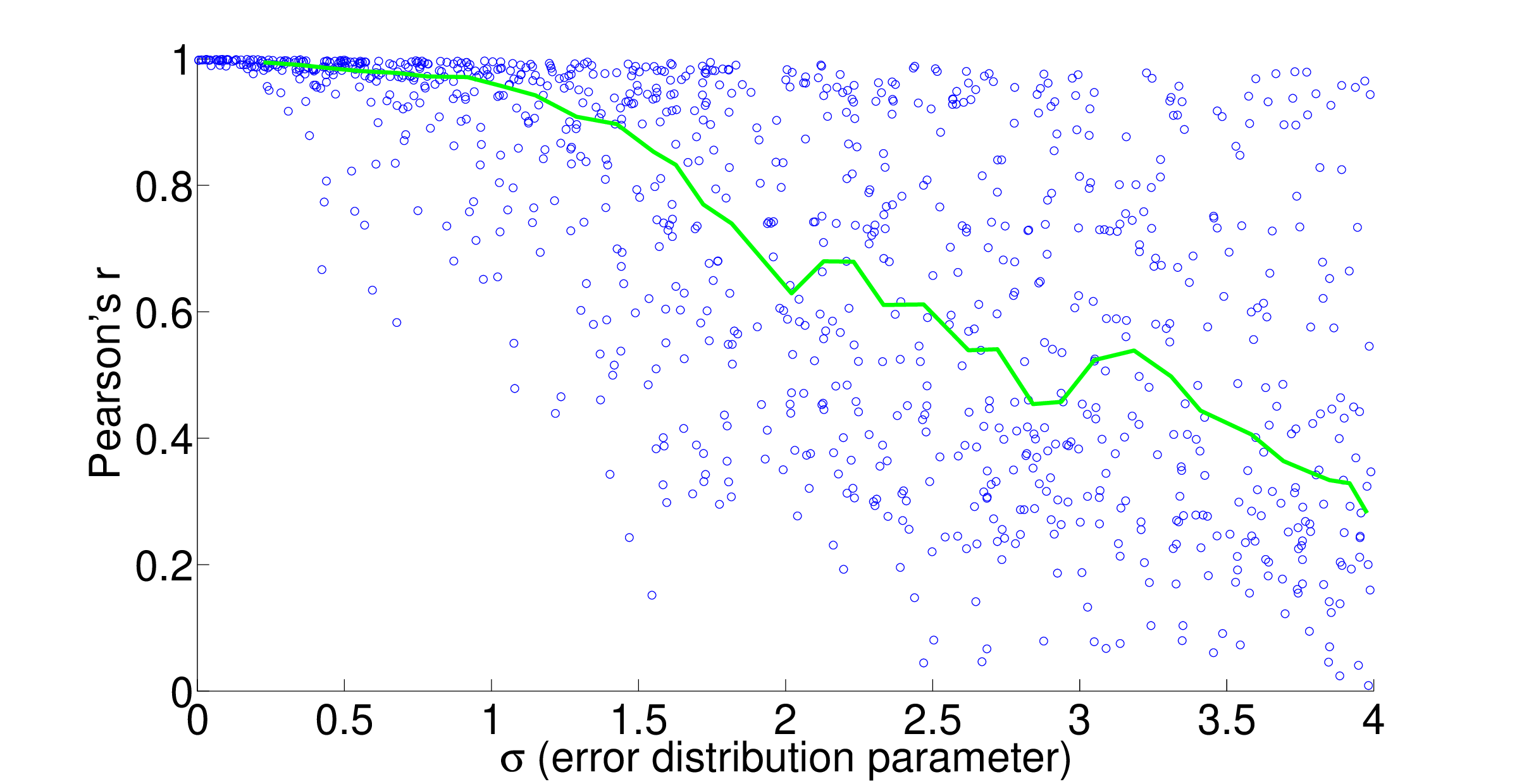}
  \caption{flux: highly constrained} \label{fig:ExpSens:B}
  \end{subfigure} 
\\
  \begin{subfigure}[b]{\expSensLen}
  \includegraphics[width=\textwidth]{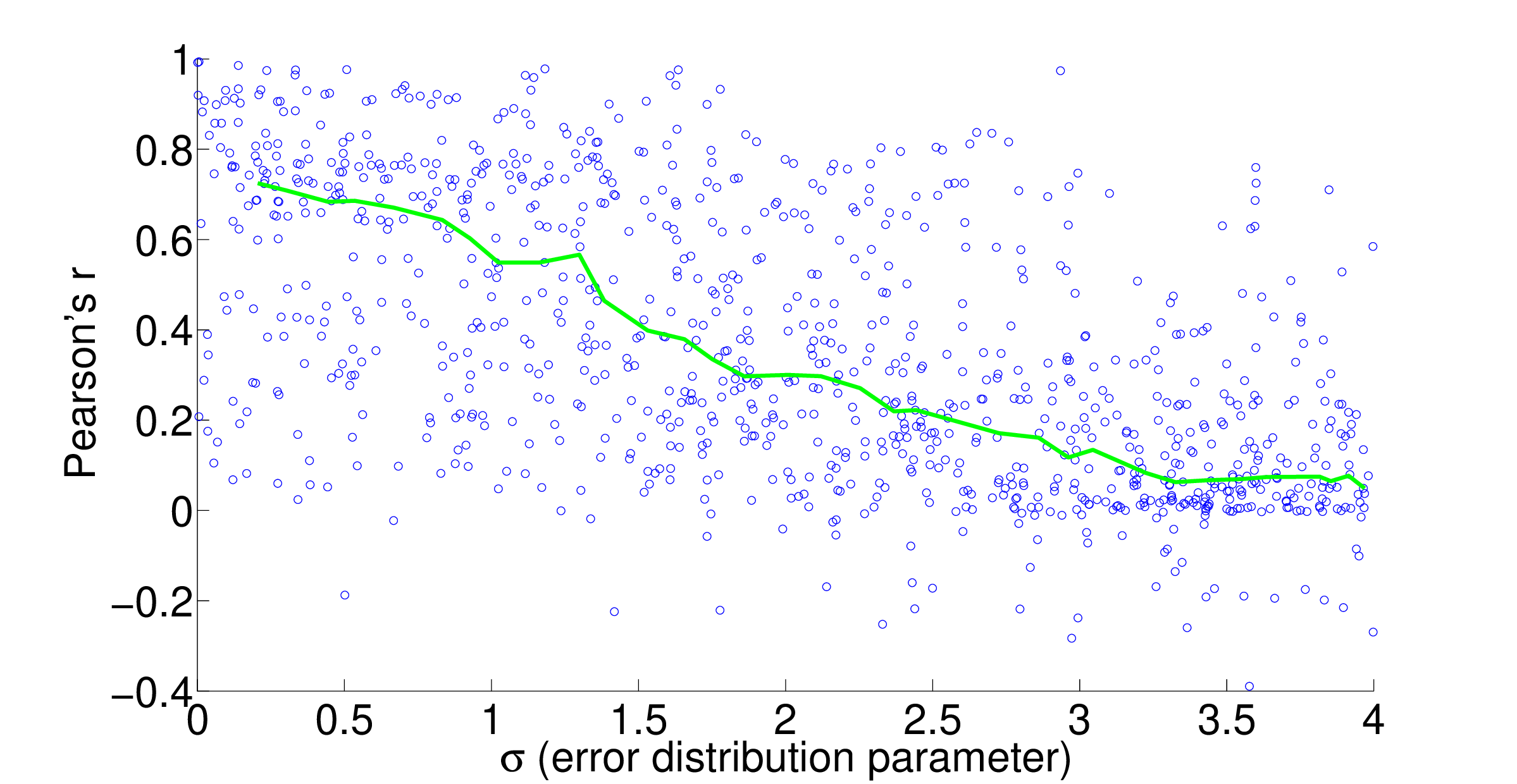}
  \caption{flux: minimally constrained} \label{fig:ExpSens:C}
  \end{subfigure} 
\\
\end{tabular}
\caption{%
Correlation of perturbed enzyme abundance vectors and flux
vectors with the associated unperturbed vector for the 
Yeast~7 model. The interval median
correlation is shown in green. Noise sampled from a multivariate
log-normal distribution with parameters $\mu=1$ and $\sigma$ (x-axis) is
multiplicatively applied to the enzyme abundance vector, and the
y-axis shows the Pearson correlation between the two vectors
\textbf{(a)}. Similar plots show correlation between flux vectors estimated
with FALCON using the same perturbed and unperturbed expression
vectors \textbf{(b-c)}.}
\label{fig:ExpSens}
\end{figure}

\subsection{Flux estimates provides information beyond enzyme complex abundance}
It is not an unreasonable hypothesis that fluxes would correlate well
with their associated complex abundances. Indeed, the general
principle needed for fitting fluxes to enzyme complex abundances is to
assume the values would be correlated in the absence of other
constraints (e.g. branch points that arise from the
stoichiometry). More specifically, it should be the case that flux is
proportional to enzyme complex abundance given ample availability of
substrate, and that this proportionality constant does not vary too
much between reactions.  There are undoubtedly many exceptions to this
rule, but it seems as though there may be some underlying evolutionary
principles for it to work in this parsimonious fashion, as has been
partly verified \citep{Bennett2009}.

Aside from the obvious benefits
of constraint-based methods also estimating fluxes for non-enzymatic
reactions, and assigning a direction for reversible enzymatic
reactions, we see that in general, our method does not predict a
strong correlation between complex abundance and flux
(\suppOrApp \Fig~\ref{fig:FluxExpCmp}). 
Recently it has been shown that many fluxes are not
under direct control of their associated enzyme expression
level \citep{Chubukov2013}, which gives experimental support to the
idea that a network-based approach, such as that
presented in this paper, may be useful in understanding how fluxes may
be constrained by expression data. \citealt{Chubukov2013} also note that enzymes
may be overexpressed in some cases, either for robustness or because
of noise in transcriptional regulation. This will not usually be a
problem in FALCON, unless entire pathways are overexpressed, which
would be unusual as it would represent a seemingly large energetic
inefficiency.

The present work doesn't attempt to use empirically
obtained kinetic parameters to estimate $V_{\max}$, but this approach
does not seem as promising in light of experimental evidence that many
reactions in central carbon metabolism tend to operate well below
$V_{\max}$ \citep{Bennett2009}. Still, a better understanding of these
phenomena may make it possible to improve flux estimation methods such
as the one presented here, or more traditional forms of MFA
\citep{Shestov2013a} by incorporating enzyme complexation and kinetic
information.

\subsection{Increasing roles for GPR rules and complex abundance estimates}
Still, complex abundance may have uses aside from being a first
step in FALCON. The method presented here for complex abundance
estimation can be used as a stand-alone method, as long as GPR
rules from a metabolic reconstruction are present. For instance, it
may not always be desirable to directly compute a flux. As an example,
the relative abundance of enzyme complexes present in secretions
from various biological tissues, such as milk or pancreatic
secretions, may still be of interest even without any intracellular
flux data. Perhaps more importantly, this approach to estimating
relative complex levels can be employed with regulatory models such as
PROM \citep{Chandrasekaran2010a} or other regulatory network models
that can estimate individual gene expression levels at time $t+1$
given the state of the model at a time $t$.

GPR rules and stoichiometry may be inaccurate or
incomplete in any given model. In fact, for the foreseeable future,
this is a given. By using the GPR and not just the stoichiometry to
estimate flux, it is possible that future work could make use of this
framework to debug not just stoichiometry as some methods currently do
(e.g.\ \citealt{Reed14112006}) , but also GPR rules.  Hope for
improved GPR rule annotation may come from many different avenues of
current research. For instance, algorithms exist for reconstructing
biological process information from large-scale datasets, and could be
tuned to aid in the annotation of GPR rules \citep{Mitra2013}. 
Flexible metabolic reconstruction pipelines such as
GLOBUS may also be extended to incorporate GPR rules into their output, and
in so doing, extend this type of modeling to many non-model organisms
\citep{Plata2012}. Another limitation that relates to lack of
biological information is that we always assume a one-to-one copy
number for each gene in a complex. Once more information on enzyme
complex structure and reaction mechanism becomes available, an
extension to the current method could make use of this information.
Even at the current level of structure, we think it is evident
that GPR rules should undergo some form of standardization;
Boolean rules without negation may not always capture the author's
intent for more complex purposes like flux fitting.

%
%

\section{Conclusion}

We have formalized and improved an existing method for estimating flux
from expression data, as well as listing detailed assumptions in
\suppOrApp Table~\ref{tab:ECAssume} that may prove useful in future
work. Although we show that expression does not correlate well with
flux, we are still essentially trying to fit fluxes to expression
levels.  The number of constraints present in metabolic models (even
the minimally constrained models) prevents a good correlation between
the two. However, as with all constraint-based models, constraints are
only part of the problem in any largely underdetermined system. We
show that gene expression can prove to be a valuable basis for forming
an objective, as opposed to methods that only use expression to
further constrain the model by creating tissue-specific or
condition-specific models \citep{Colijn2009,Shlomi2008,Becker2008,Gowen2010}.

For better curated models, the approach described immediately finds
use for understanding metabolism, as well as being a scaffold to find
problems for existing GPR rules, and more broadly the GPR formalism itself.
The present results and avenues for future improvement
show that there is much promise for using expression to estimate
fluxes, and that it can already be a useful tool for performing flux
estimation and analysis.

\section{Acknowledgments}
We thank Neil Swainston for his previous work on these research issues
and many helpful discussions about considerations in various genome
scale models of metabolism.  We also thank Michael Stillman for
general discussions about modeling and for reading the paper. Finally
we thank Alex Shestov, Lei Huang, and Eli Bogart for many helpful
discussions about metabolism.

BEB and YW are supported by the Tri-Institutional Training Program in
Computational Biology and Medicine. BEB and ZG are thankful for
support from NSF grant MCB-1243588.  KS is grateful for the financial
support of the EU FP7 project BioPreDyn (grant 289434). CRM
acknowledges support from NSF Grant IOS-1127017. JWL thanks the NIH
for support through grants R00 CA168997 and R01 AI110613.

%% file: common/falcon_appendix.tex
\renewcommand{\figurename}{\textbf{\suppOrApp \Fig}}
\renewcommand{\tablename}{\textbf{\suppOrApp Table}}

\begin{figure}
\section{Supporting figures}
\centering
\begin{tabular}{cc}
  \begin{subfigure}[b]{0.5\textwidth}
  \includegraphics[width=\textwidth, trim=9cm 0cm 9cm 0cm, clip=true]
  {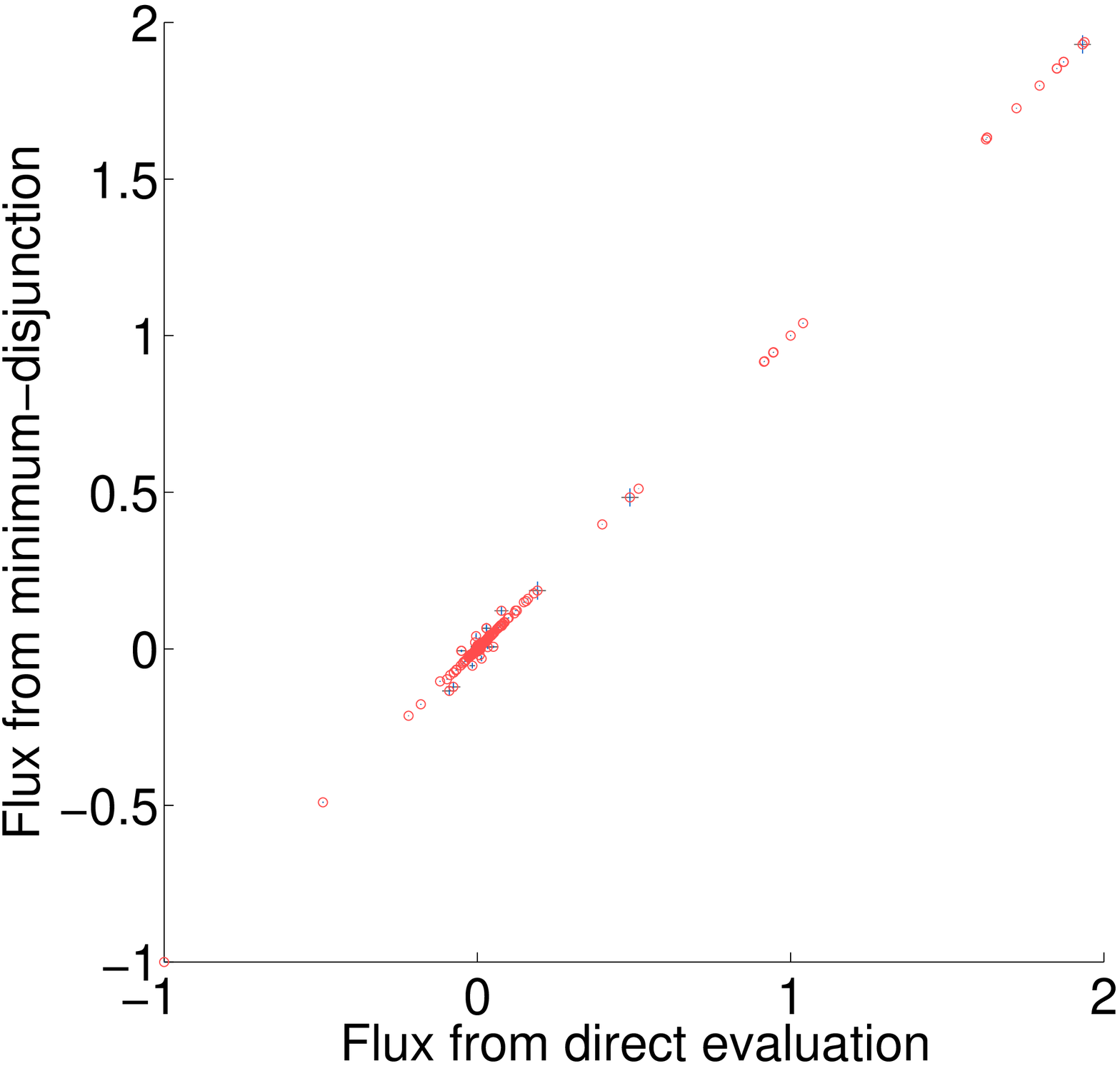}
  \caption{highly constrained} \label{fig:EnzAbundEval:A}
  \end{subfigure}
&
  \begin{subfigure}[b]{0.5\textwidth}
  \includegraphics[width=\textwidth, trim=9cm 0cm 9cm 0cm, clip=true]
  {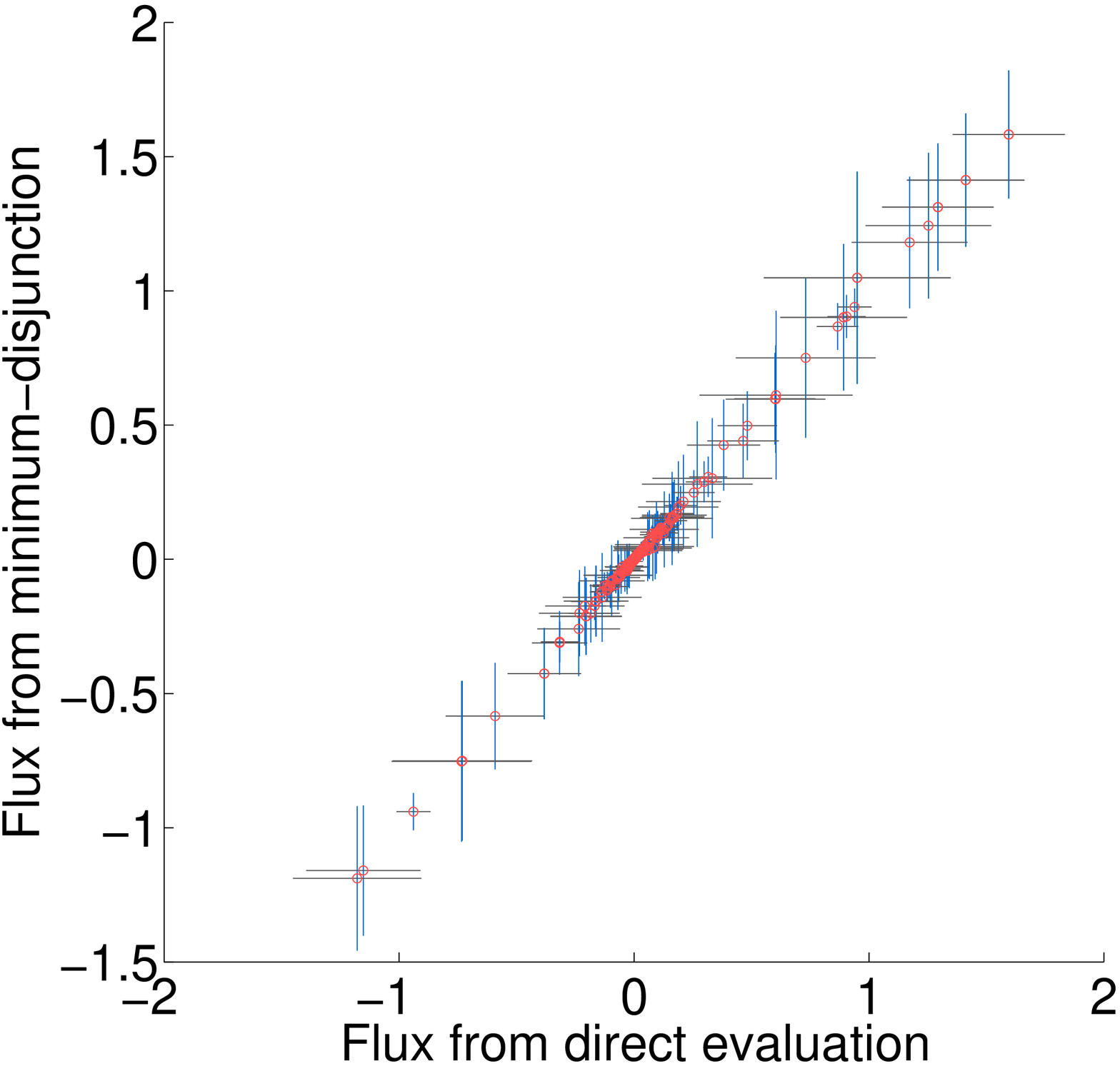}
  \caption{minimally constrained} \label{fig:EnzAbundEval:B}
  \end{subfigure} 
\\
  \begin{subfigure}[b]{0.5\textwidth}
  \includegraphics[width=\textwidth, trim=9cm 0cm 9cm 0cm, clip=true]
  {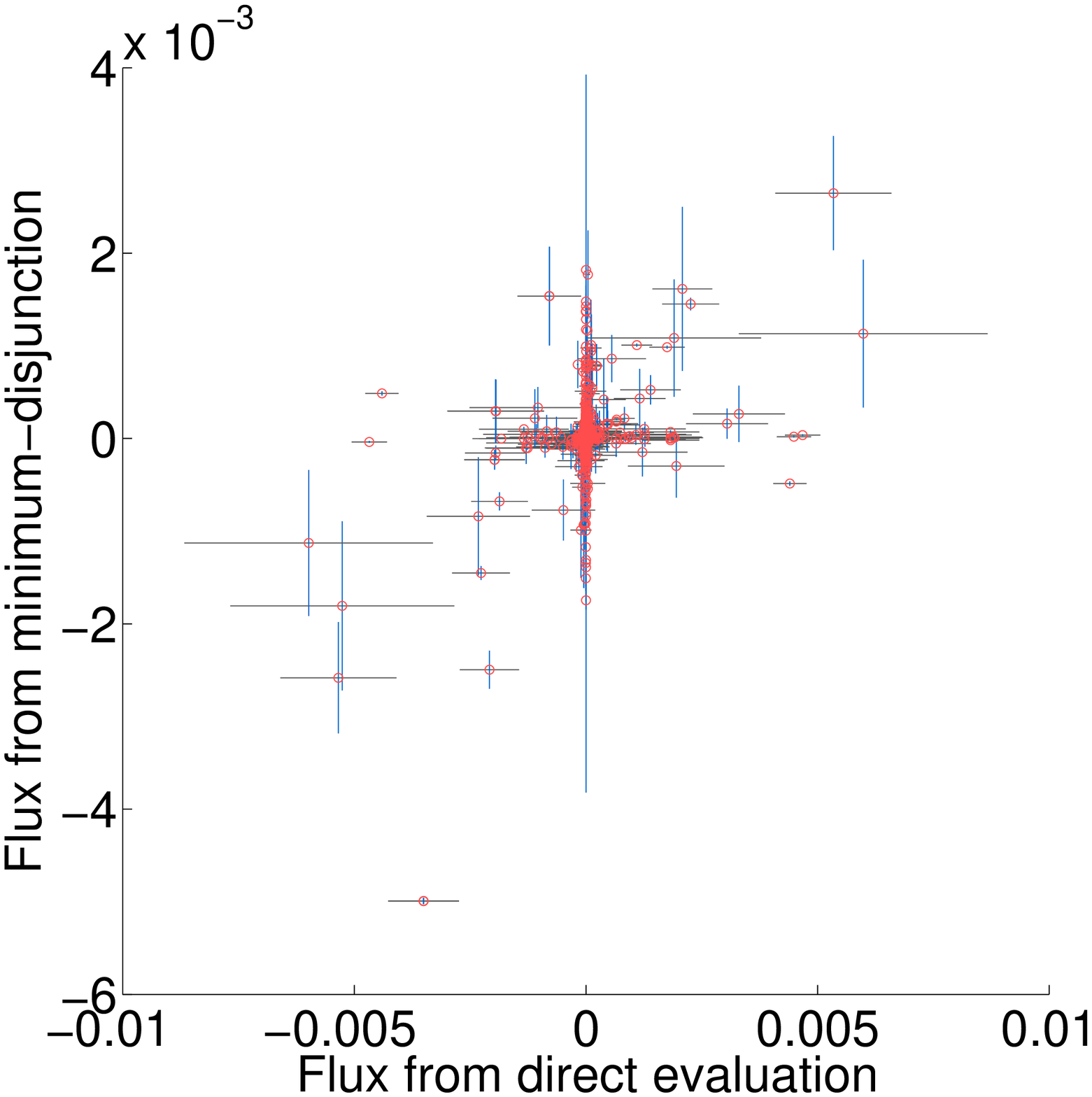}
  \caption{highly constrained} \label{fig:EnzAbundEval:C}
  \end{subfigure}
&
  \begin{subfigure}[b]{0.5\textwidth}
  \includegraphics[width=\textwidth, trim=9cm 0cm 9cm 0cm, clip=true]
  {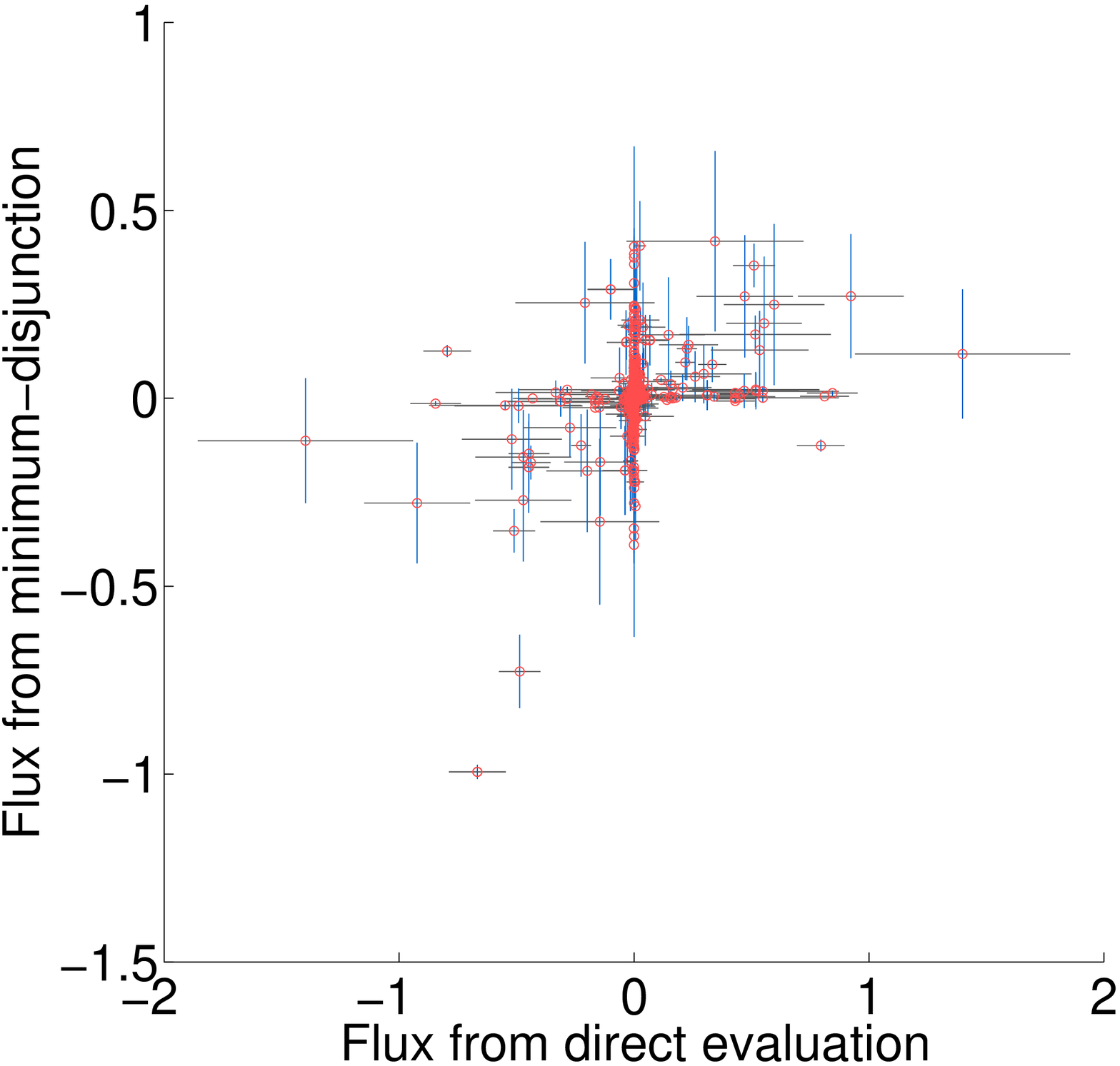}
  \caption{minimally constrained} \label{fig:EnzAbundEval:D}
  \end{subfigure} 
\\
\end{tabular}
\caption{Comparison of fluxes when FALCON is run with enzyme abundance
calculated by direct evaluation (x-axis) and the minimum disjunction
algorithm (y-axis); error bars with length equal to one standard
deviation are shown for both approaches as a result of alternative
solutions in FALCON. Yeast was evaluated with default (highly)
constrained \textbf{(a)} and minimally constrained \textbf{(b)}
models, and no strong difference between direct evaluation or or the
minimum disjunction method is observed in either case. However, for
human models with a highly constrained reaction set (RPMI media,
CORE-sign, and enzymatic direction) \textbf{(c)} and default
constraints \textbf{(d)}, we see there is a large amount of variation
between the two evaluation techniques. In the human cases, two
outliers were not shown that correspond to a single large flux cycle
(`release of B12 by simple diffusion' and `transport of
Adenosylcobalamin into the intestine').}
\label{fig:EnzAbundEval}
\end{figure}
\FloatBarrier

\begin{figure}[!htb]
\begin{tabular}{cc}
  \begin{subfigure}[b]{0.48\textwidth}
  \includegraphics[width=\textwidth]{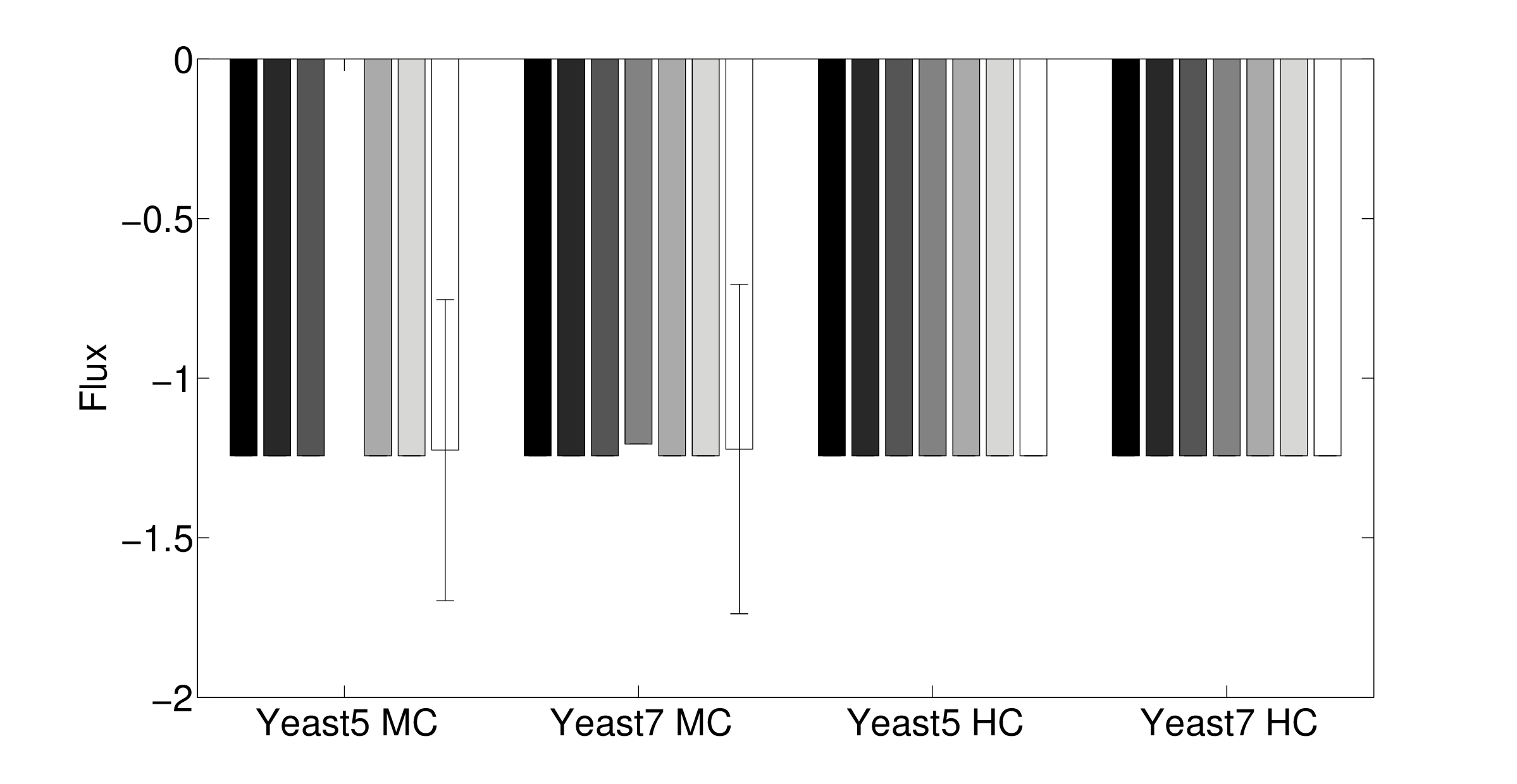}
  \caption{glucose}
  \end{subfigure}
&
  \begin{subfigure}[b]{0.48\textwidth}
  \raisebox{0.3\height}{\includegraphics[scale=0.4,center]{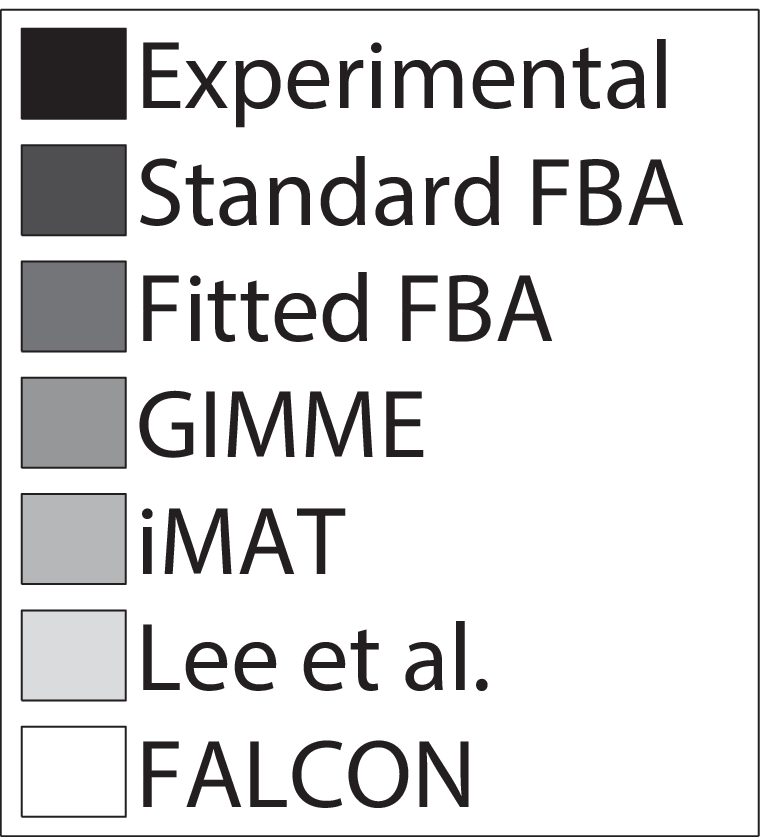}}
  \end{subfigure} 
\\
  \begin{subfigure}[b]{0.48\textwidth}
  \includegraphics[width=\textwidth]{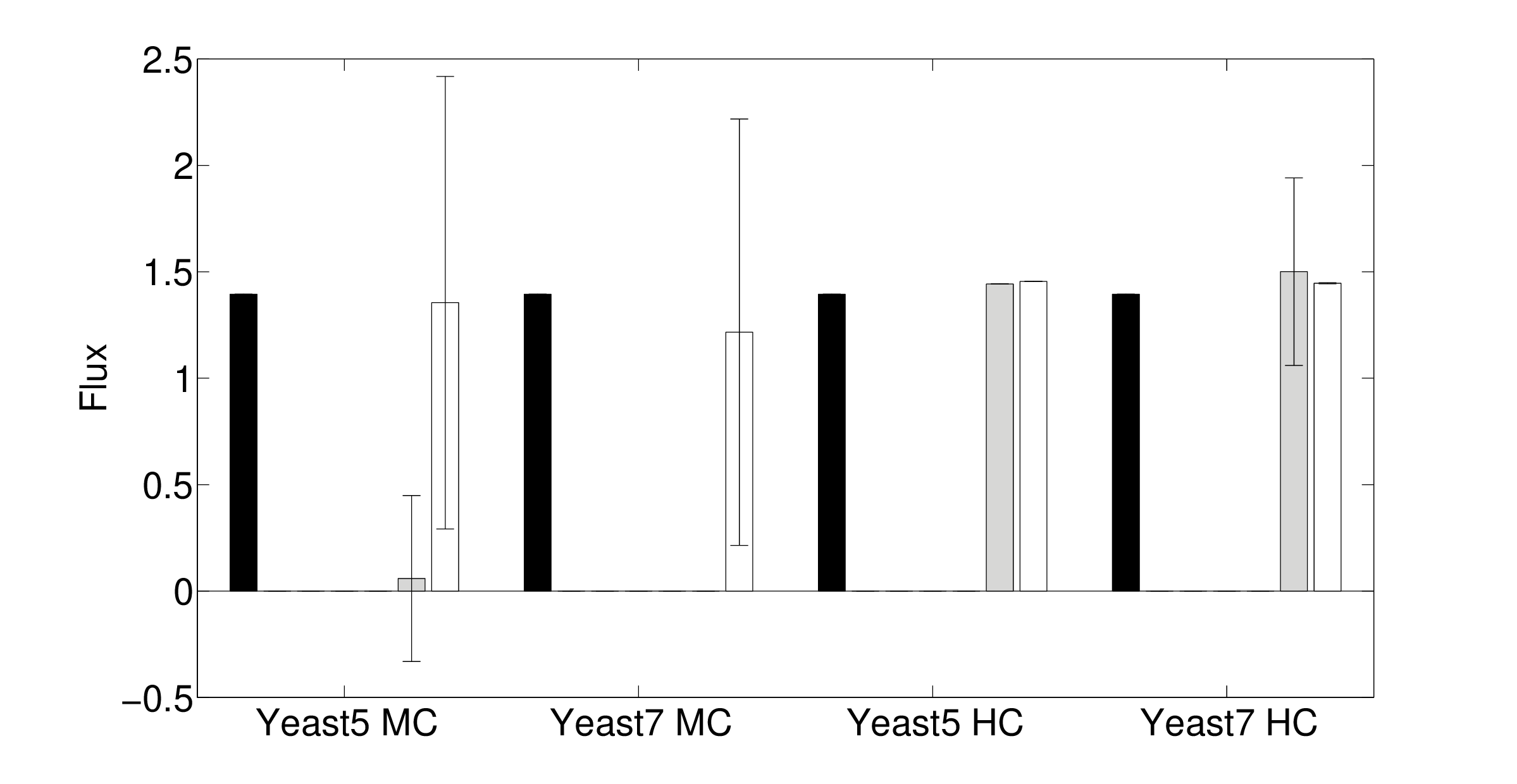}
  \caption{ethanol}
  \end{subfigure} 
&
  \begin{subfigure}[b]{0.48\textwidth}
  \includegraphics[width=\textwidth]{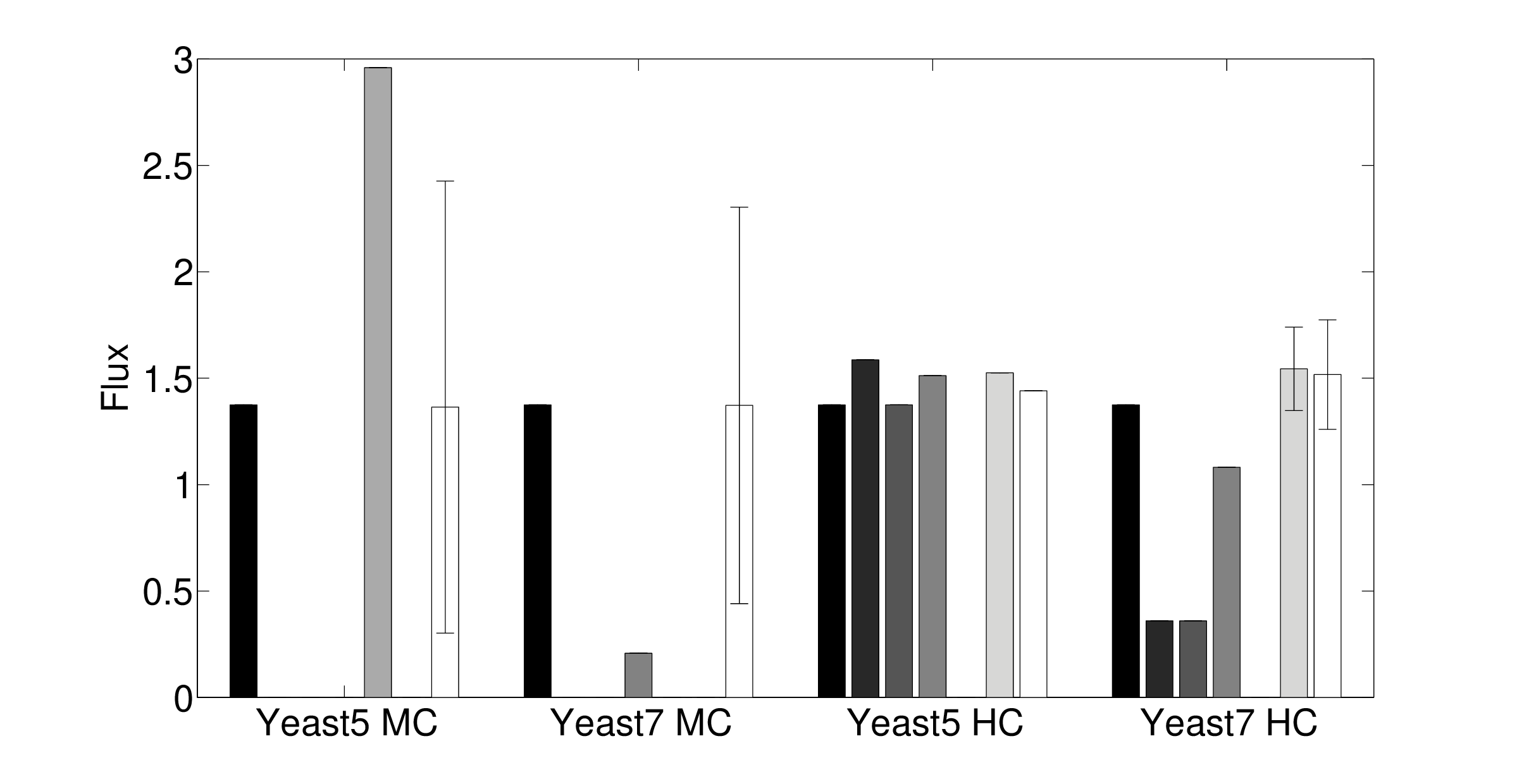}
  \caption{CO$_2$}
  \end{subfigure} 
\\
  \begin{subfigure}[b]{0.48\textwidth}
  \includegraphics[width=\textwidth]{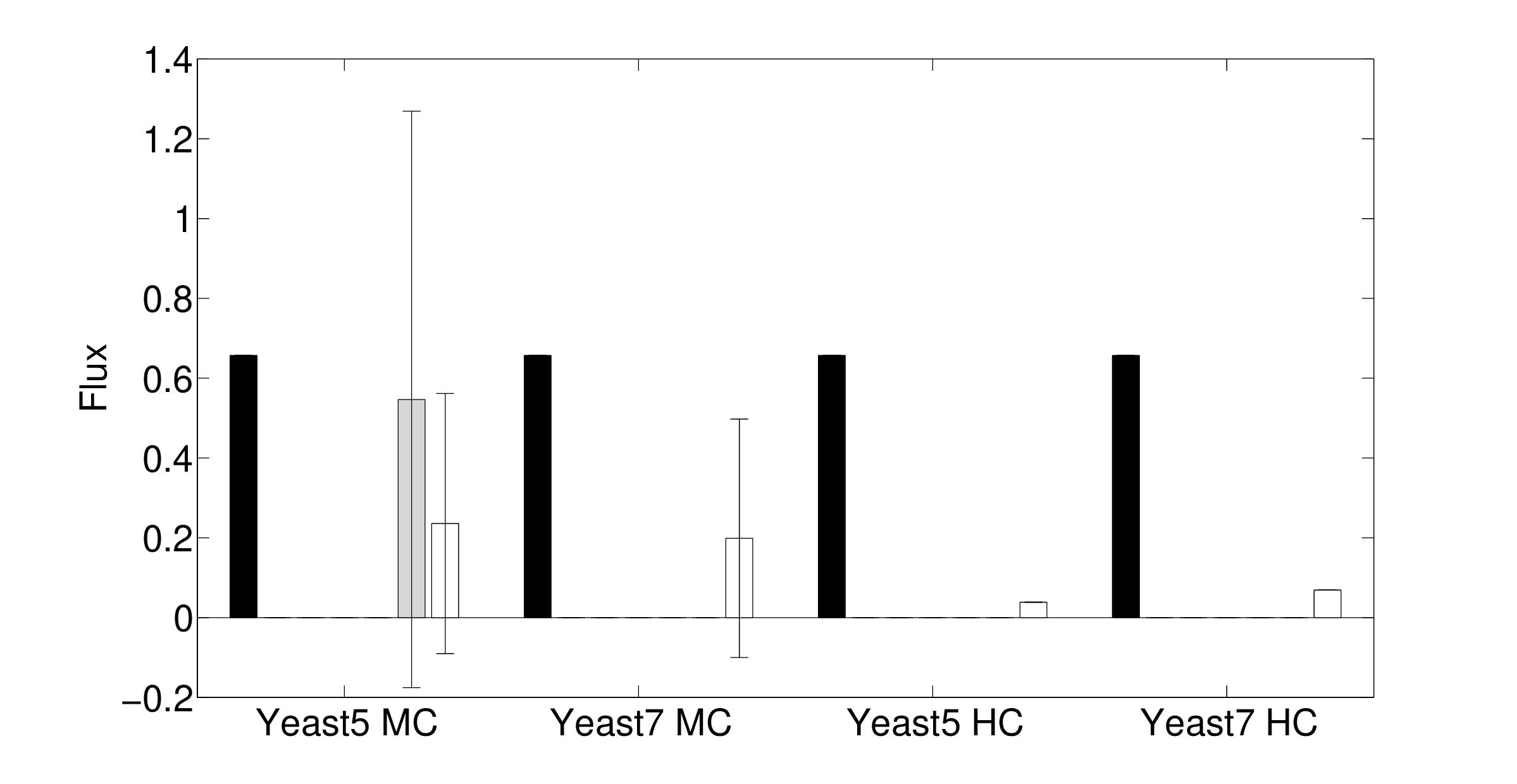}
  \caption{glycerol}
  \end{subfigure} 
&
  \begin{subfigure}[b]{0.48\textwidth}
  \includegraphics[width=\textwidth]{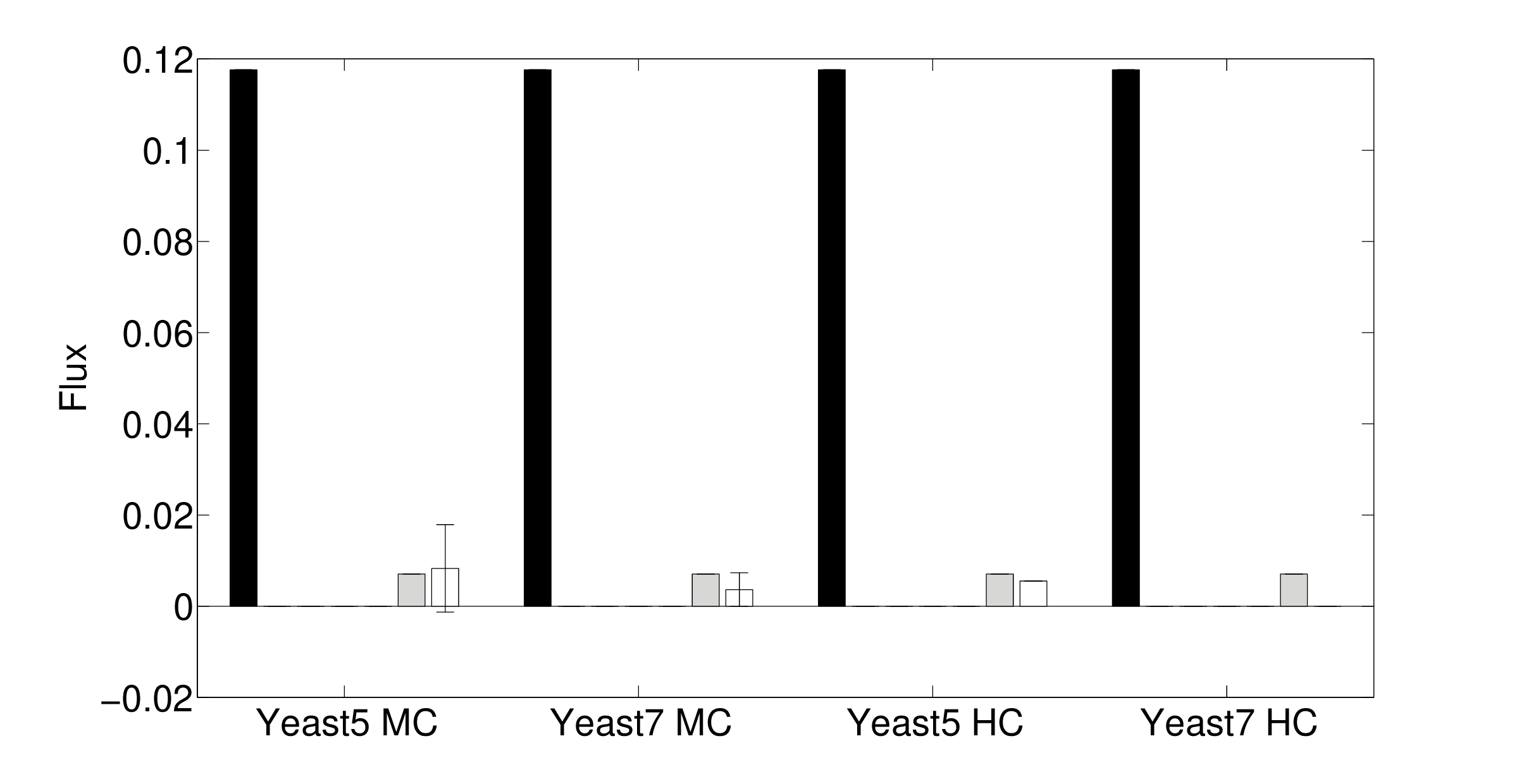}
  \caption{acetate}
  \end{subfigure} 
\\
\end{tabular}
\caption{
Shown are flux predictions using a number of methods and four
different models (Yeast~5 MC and Yeast~7 MC are minimally constrained
Yeast~5 and Yeast~7; Yeast~5 HC and Yeast~7 HC are highly constrained Yeast~5
and Yeast~7). Error bars are shown for the Lee et al. method and for
FALCON, where one side of the error bar corresponds to a standard
deviation. Note that there can be no variation for glucose in the
former case since glucose flux is fixed as part of the method. FALCON
performs very well for large fluxes \textbf{(a-c)}, and is also the best
performer in general for the next largest flux, glycerol \textbf{(d)}. It also
has sporadic success for smaller fluxes, but all methods seem to have
trouble with the smallest fluxes (e.g. \textbf{e}). Note that fluxes are drawn
in log scale (specifically a flux $v$ is drawn as $\sgn{v} \log_{10}{\left( 
1 + v\right)} $). Similar results are obtainable for the 85\% maximum growth
condition.}
\label{fig:FluxBars}
\end{figure}
\FloatBarrier

\begin{figure}[!htb]
\centering
\begin{tabular}{cc}
  \begin{subfigure}[b]{0.5\textwidth}
  \includegraphics[width=\textwidth]{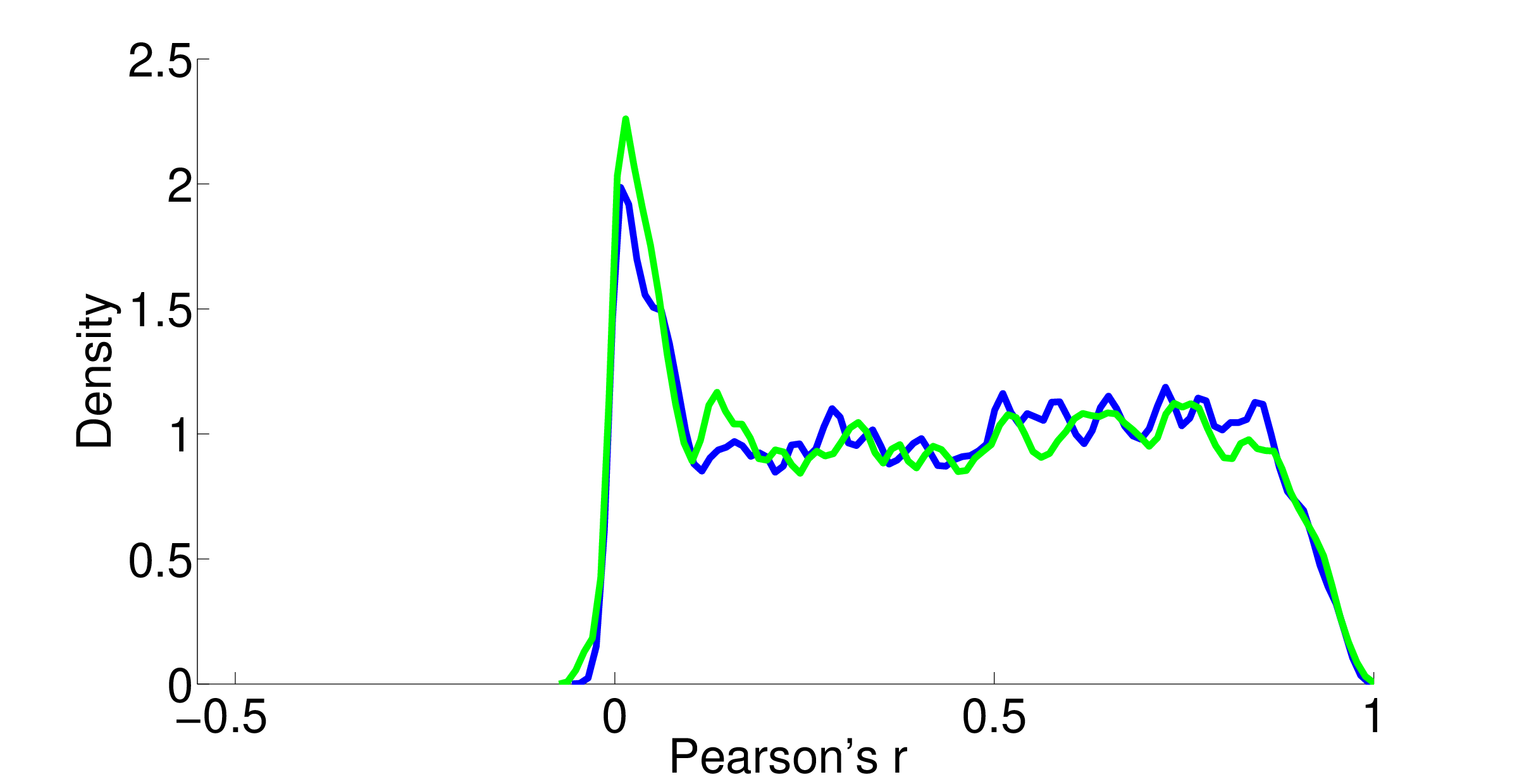}
  \caption{highly constrained} \label{fig:YpermCorrSup:A}
  \end{subfigure}
&
  \begin{subfigure}[b]{0.5\textwidth}
  \includegraphics[width=\textwidth]{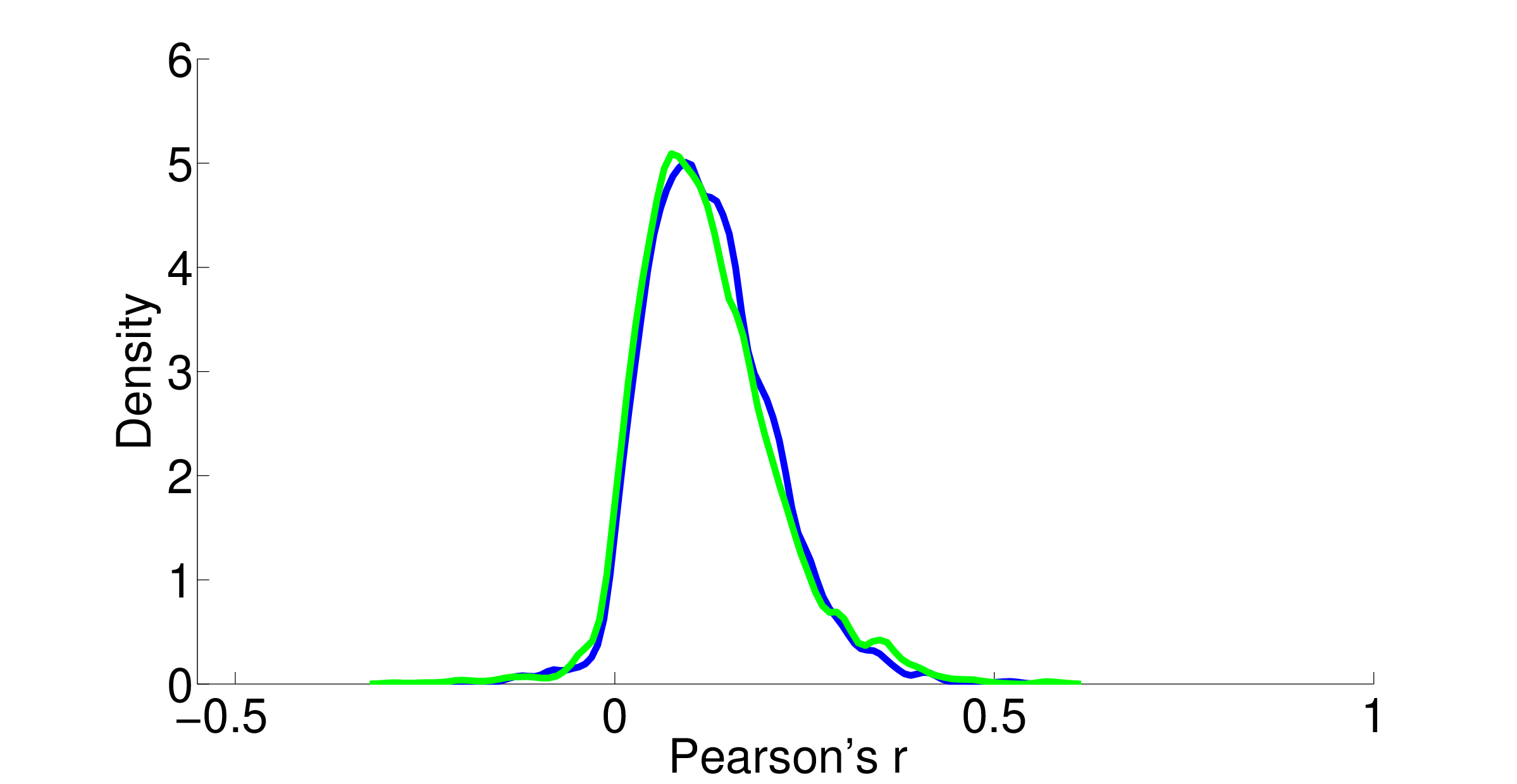}
  \caption{minimally constrained} \label{fig:YpermCorrSup:B}
  \end{subfigure} 
\\
  \begin{subfigure}[b]{0.5\textwidth}
  \includegraphics[width=\textwidth]{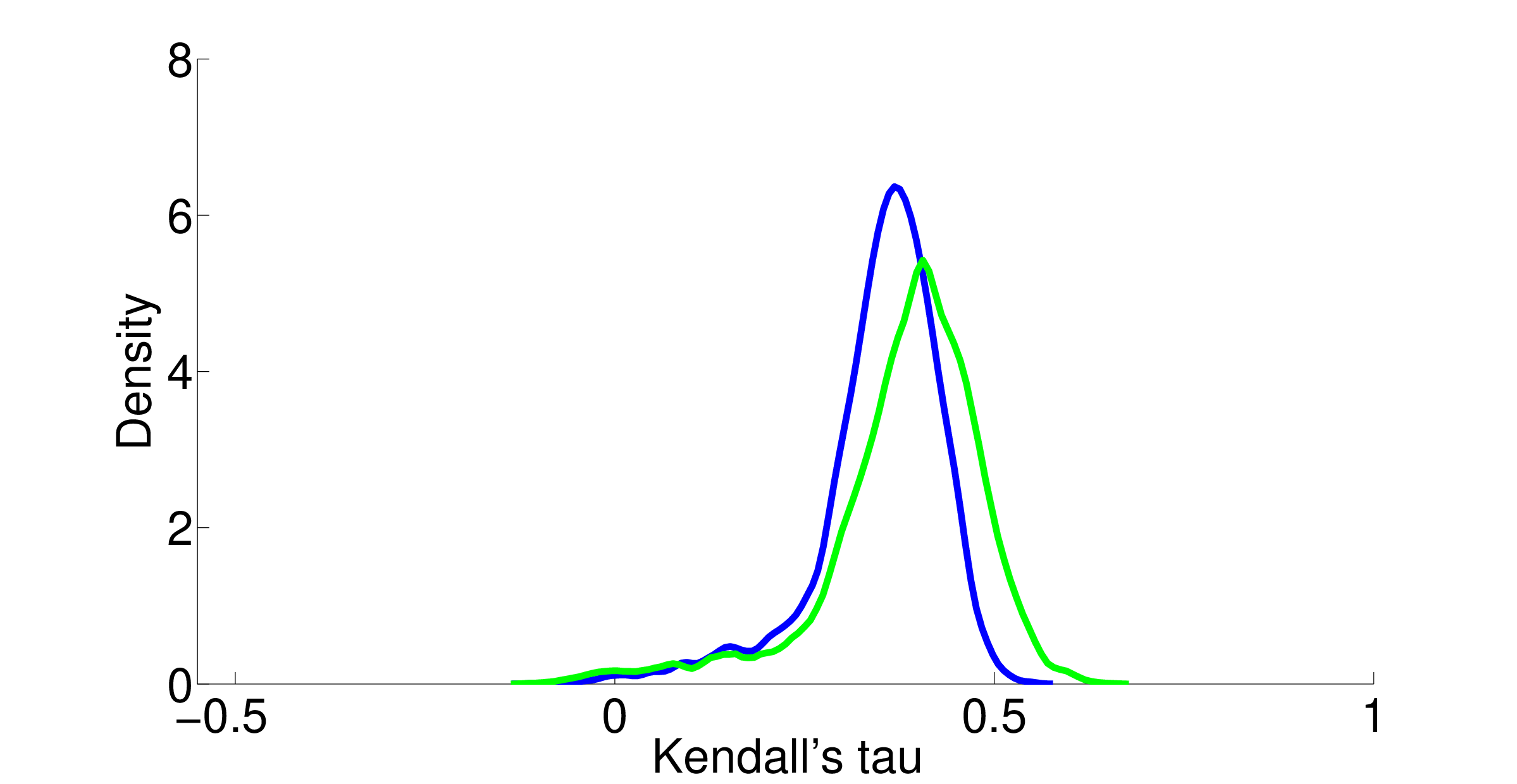}
  \caption{highly constrained} \label{fig:YpermCorrSup:C}
  \end{subfigure}
&
  \begin{subfigure}[b]{0.5\textwidth}
  \includegraphics[width=\textwidth]{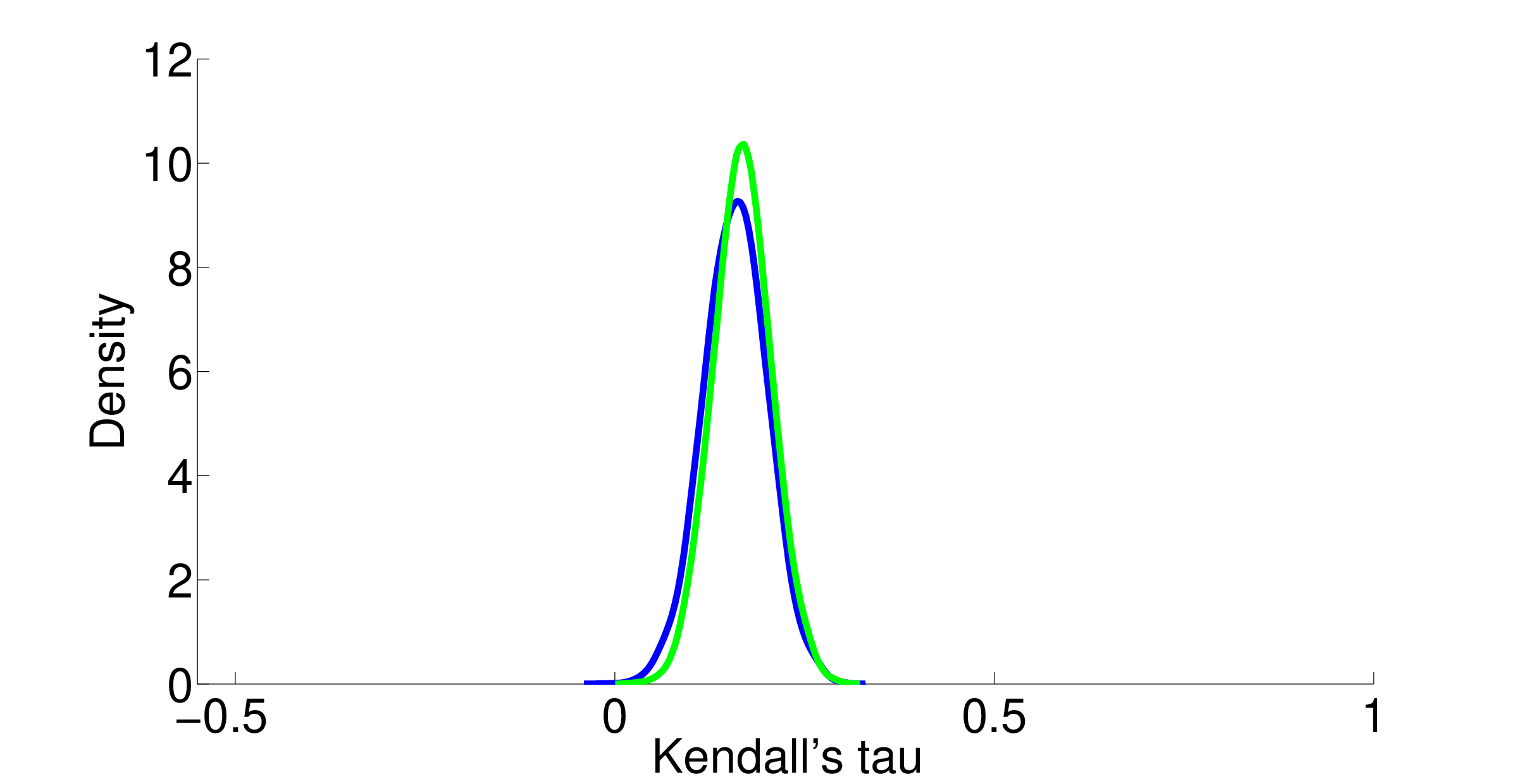}
  \caption{minimally constrained} \label{fig:YpermCorrSup:D}
  \end{subfigure} 
\\
\end{tabular}
\caption{Kernel-smoothed PDFs are drawn for correlations between
the entire flux vector estimated by FALCON on permuted and unpermuted
data. Stability and correlation are effected by constraints, as there
are differences between the minimally constrained \textbf{(b, d)} and
highly constrained \textbf{(a, c)} Yeast~7 models. 5,000 permutation
replicates were performed in all cases.}
\label{fig:YpermCorrSup}
\end{figure}
\FloatBarrier

\begin{figure}[!htb]
\begin{tabular}{cc}
  \begin{subfigure}[b]{0.5\textwidth}
  \includegraphics[width=\textwidth]{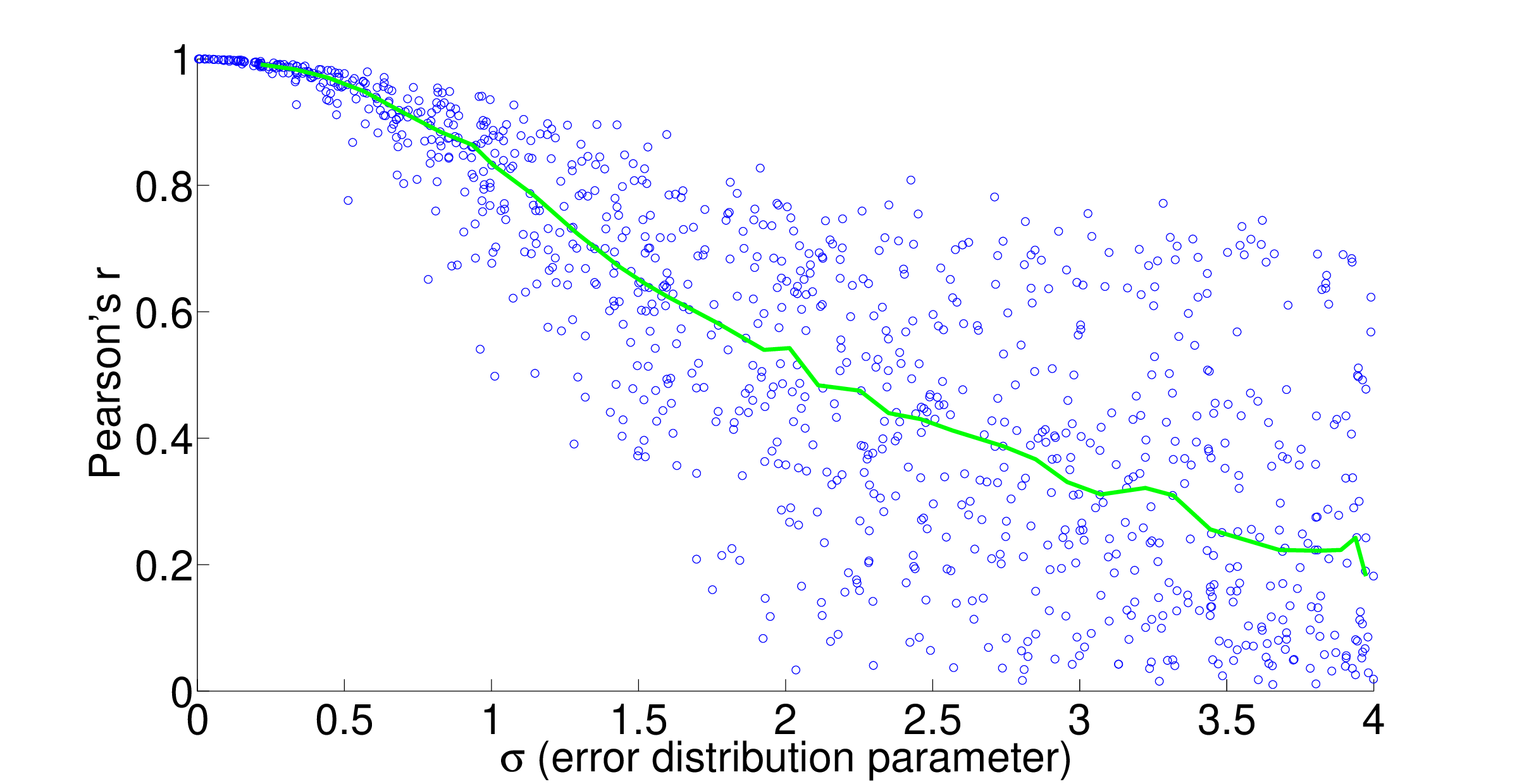}
  \caption{} \label{fig:ExpSensRec2:A}
  \end{subfigure}
&
  \begin{subfigure}[b]{0.5\textwidth}
  \includegraphics[width=\textwidth]{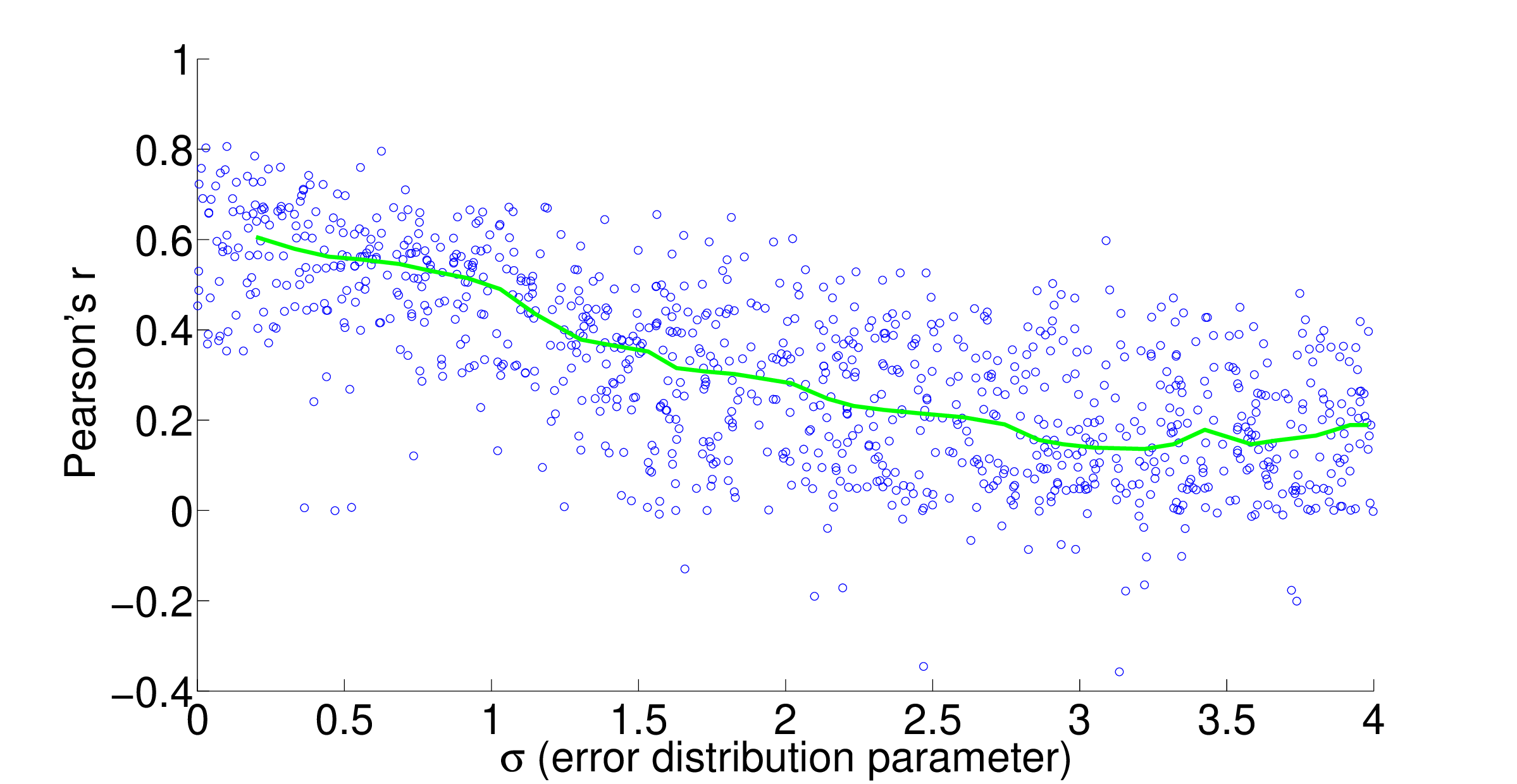}
  \caption{} \label{fig:ExpSensRec2:B}
  \end{subfigure} 
\\
  \begin{subfigure}[b]{0.5\textwidth}
  \includegraphics[width=\textwidth]{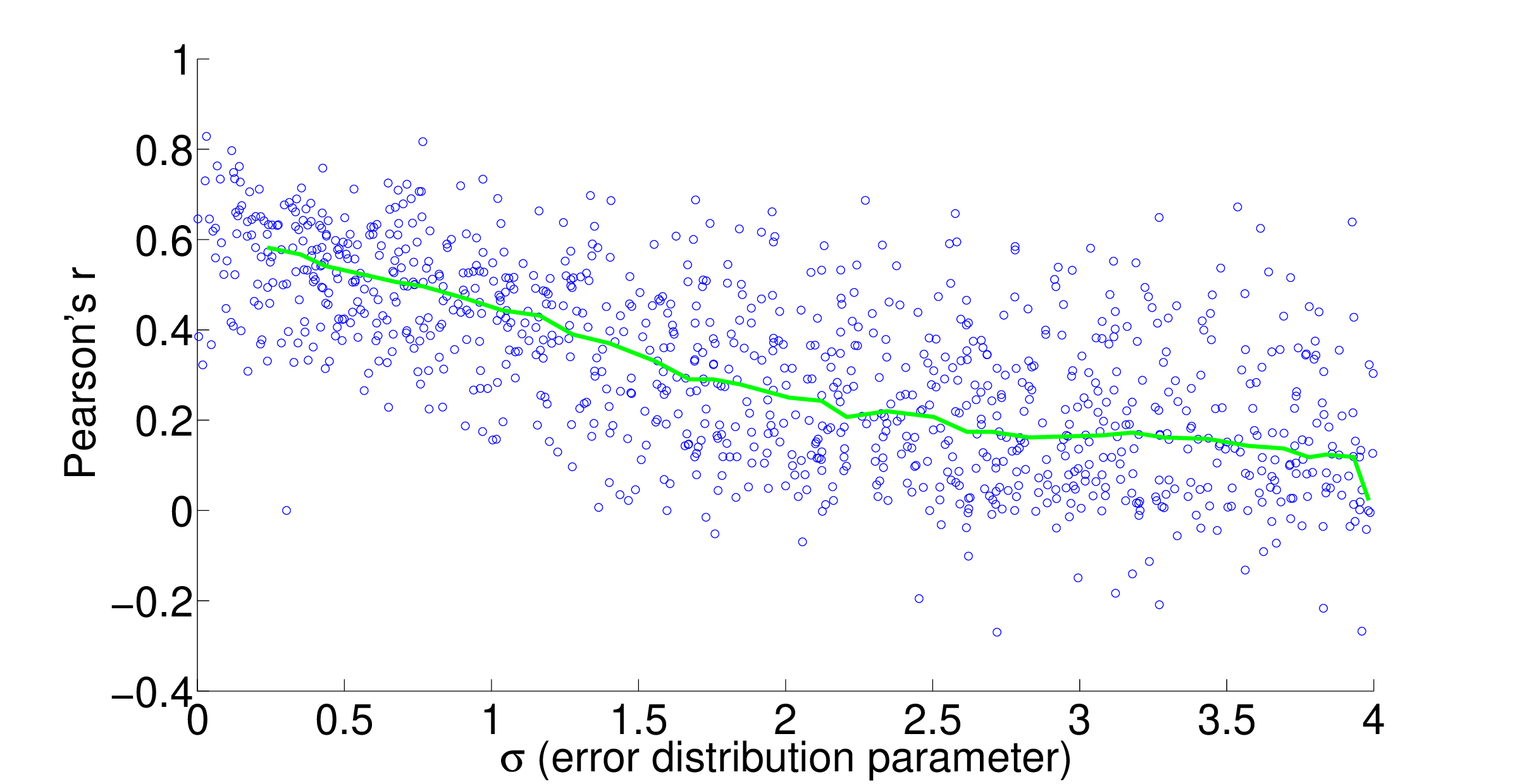}
  \caption{} \label{fig:ExpSensRec2:C}
  \end{subfigure} 
&
  \begin{subfigure}[b]{0.5\textwidth}
  \includegraphics[width=\textwidth]{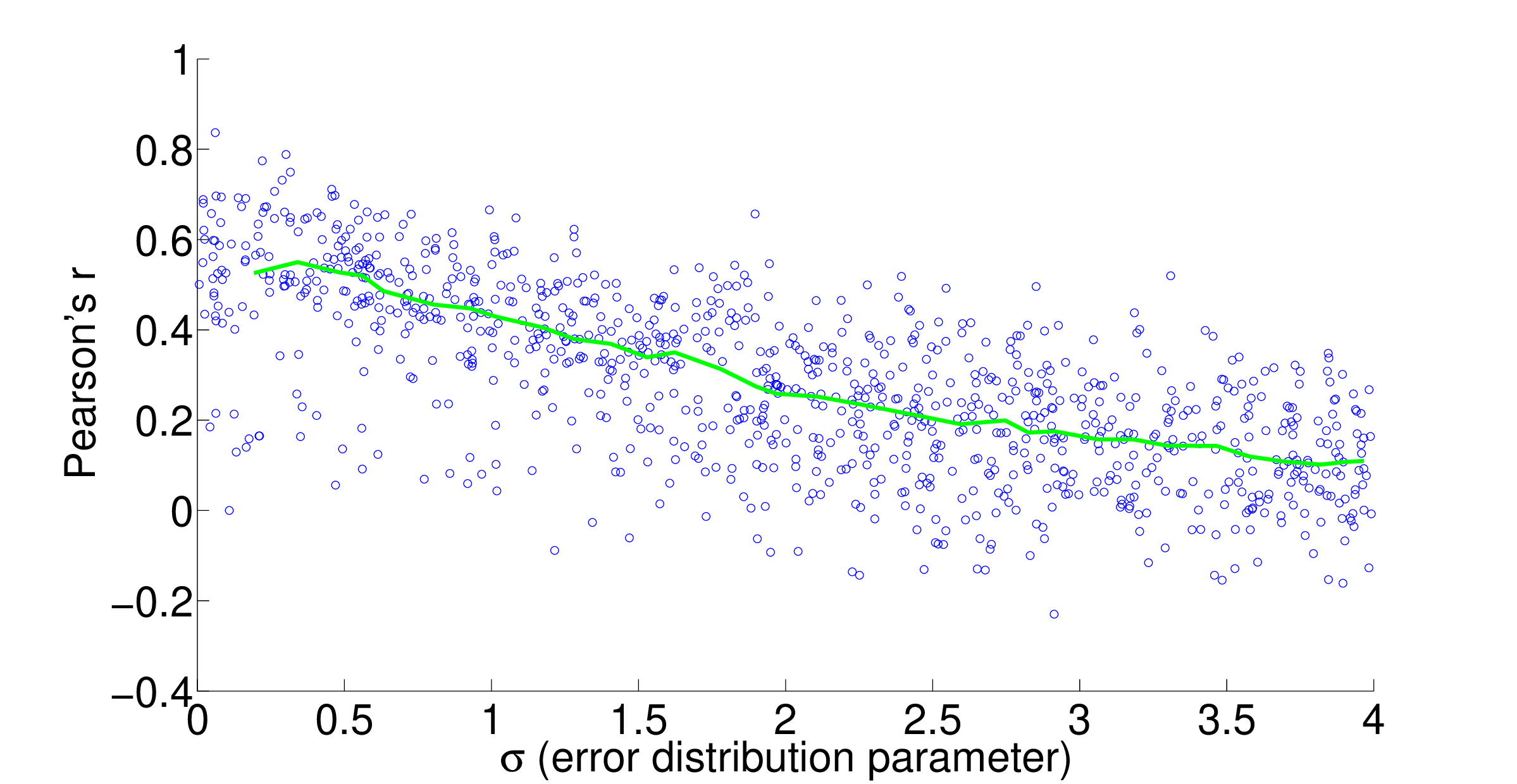}
  \caption{} \label{fig:ExpSensRec2:D}
  \end{subfigure} 
\\
  \begin{subfigure}[b]{0.5\textwidth}
  \includegraphics[width=\textwidth]{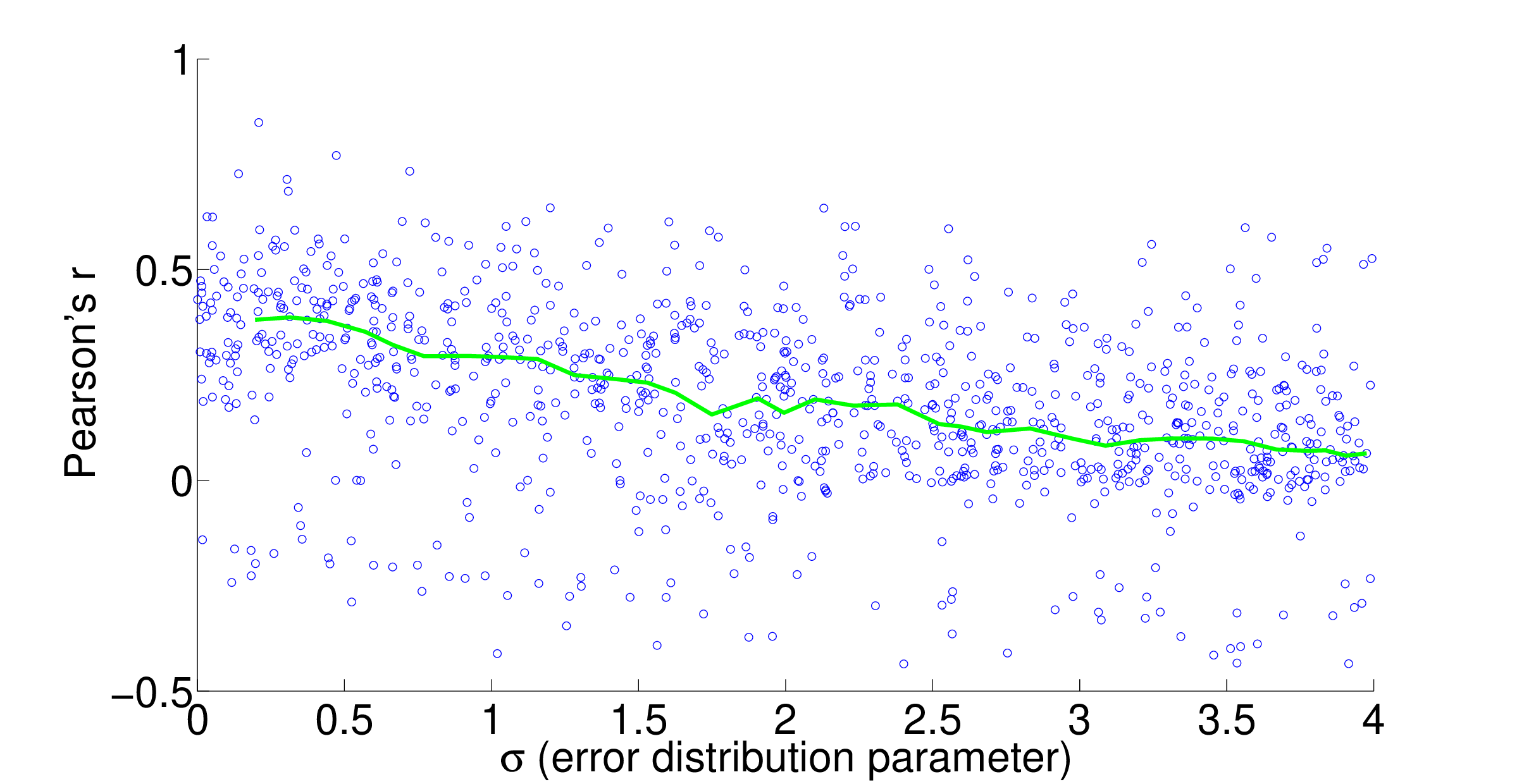}
  \caption{} \label{fig:ExpSensRec2:E}
  \end{subfigure} 
&
  \begin{subfigure}[b]{0.5\textwidth}
  \includegraphics[width=\textwidth]{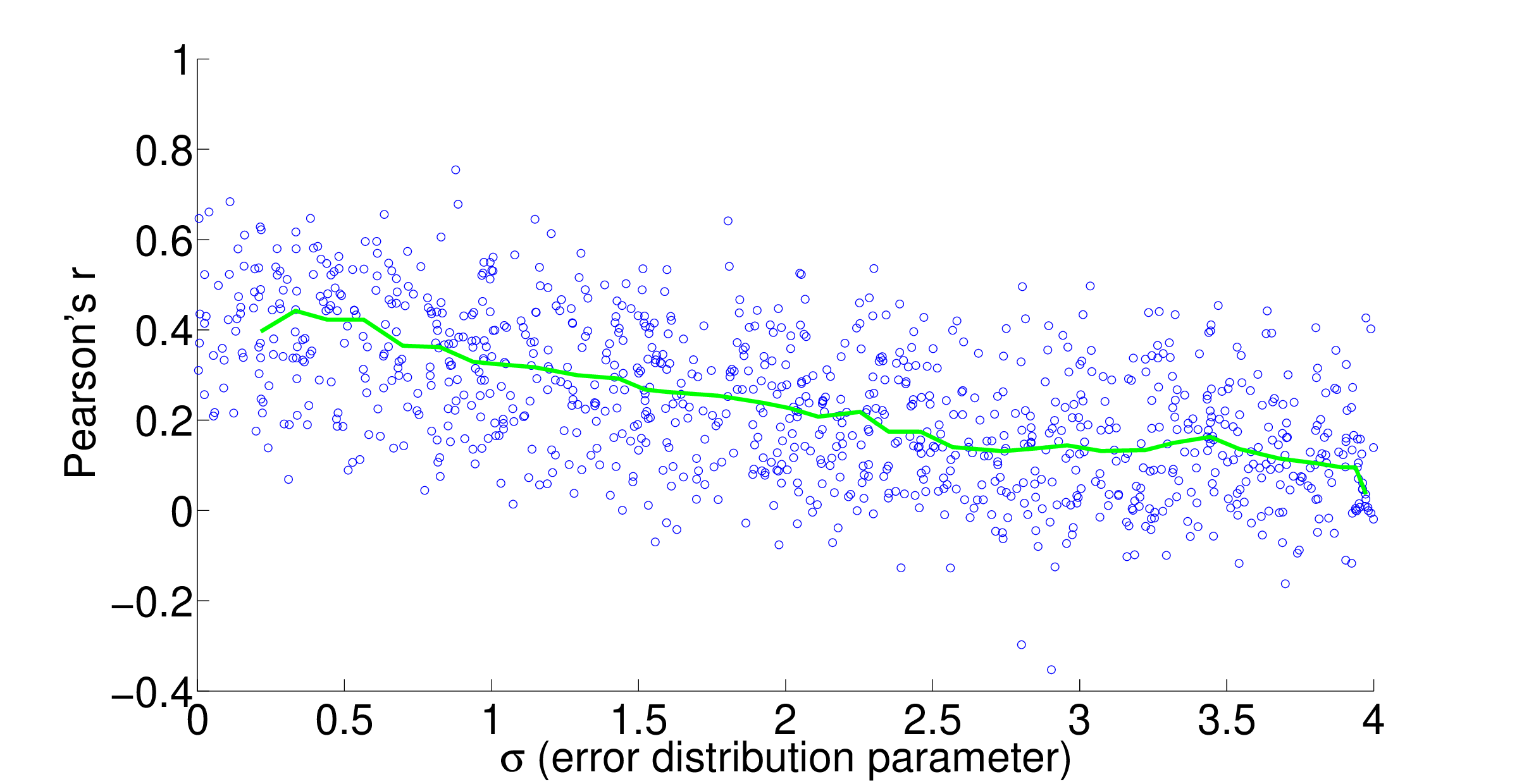}
  \caption{} \label{fig:ExpSensRec2:F}
  \end{subfigure} 
\\
\end{tabular}
\caption{ These figures are generated in the same way as those in
\Fig~\ref{fig:ExpSens}, but for Human Recon~2 instead of Yeast~7, where
\textbf{(a)} again shows how noise applied to an expression vector (x-axis)
influences the Pearson correlation between the perturbed and unperturbed
expression vectors (y-axis). We
used several different constraint sets based on experimental media and
exometabolic flux data in the NCI-60 cell lines \protect\citep{Jain2012}. 
These constraints were applied cumulatively, and are
listed in the order of most constrained \textbf{(b)} to least
constrained \textbf{(f)}. Included are default Recon~2 constraints
\textbf{(f)}, RPMI media constraints (\textbf{e}; function
\texttt{constrainCoReMinMaxSign}; 556 constraints), exometabolic
fluxes with a common sign across all cell lines and replicates
(\textbf{d}; function \texttt{constrainCoReMinMaxSign}; 567 cumulative
constraints), enzymatic reaction directionality constraints from a
linear MoMA fitting on the exometabolic flux data that agree across
all NCI-60 cell lines (\textbf{c};
\texttt{constrainImputed\-Internal}; 593 cumulative constraints), and
the same again considering all reactions instead of only enzymatic
reactions (\textbf{b}; 618 cumulative constraints).}
\label{fig:ExpSensRec2}
\end{figure}
\FloatBarrier

\begin{figure}[!htb]
\begin{tabular}{cc}
  \begin{subfigure}[b]{0.5\textwidth}   
  \includegraphics[width=\textwidth, trim=9cm 1.2cm 9cm 1cm, clip=true]
    {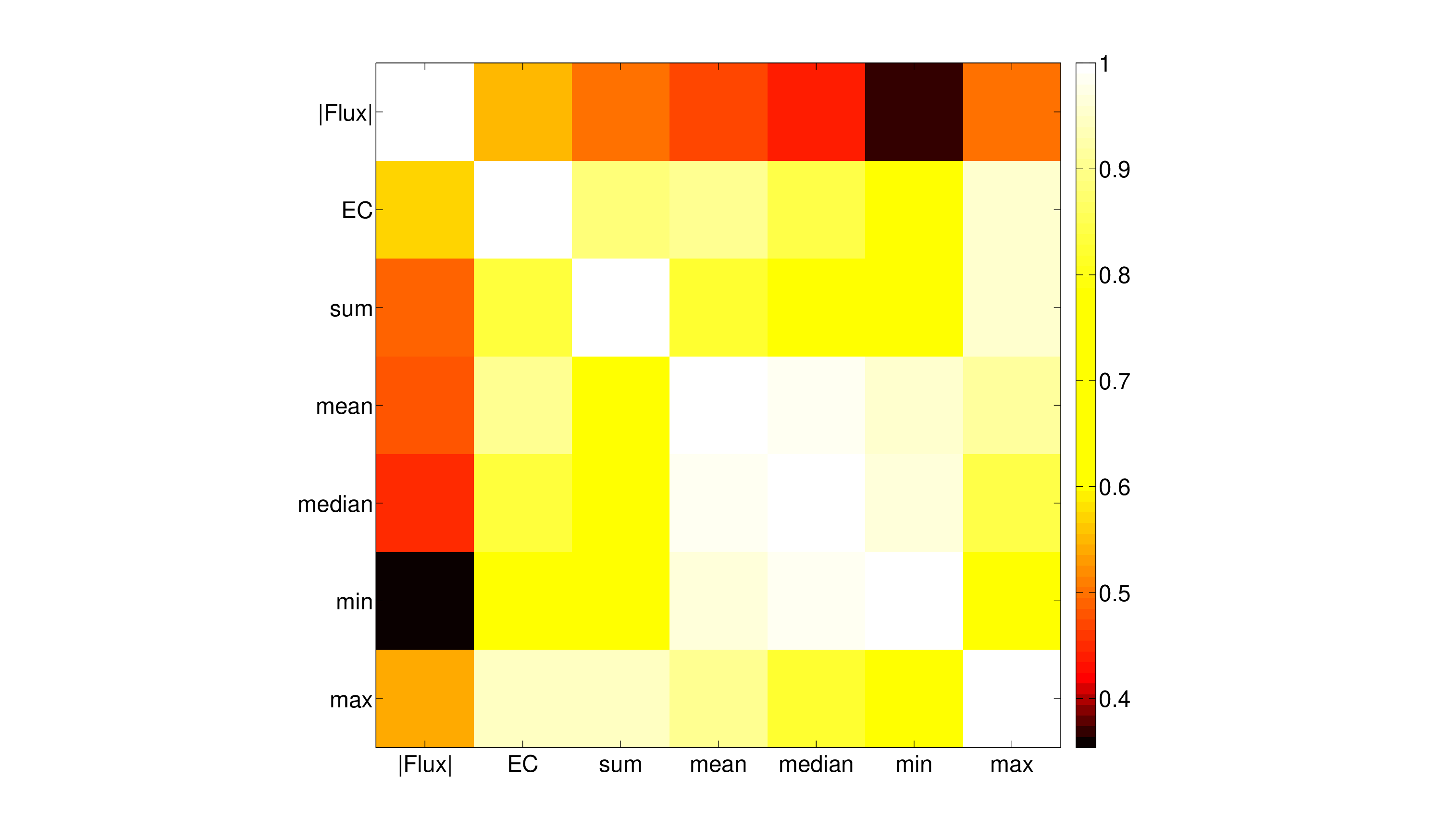}
  \caption{yeast} \label{fig:FluxExpCmp:A}
  \end{subfigure}
&
  \begin{subfigure}[b]{0.5\textwidth}
  \includegraphics[width=\textwidth, trim=9cm 1.2cm 9cm 1cm, clip=true]
    {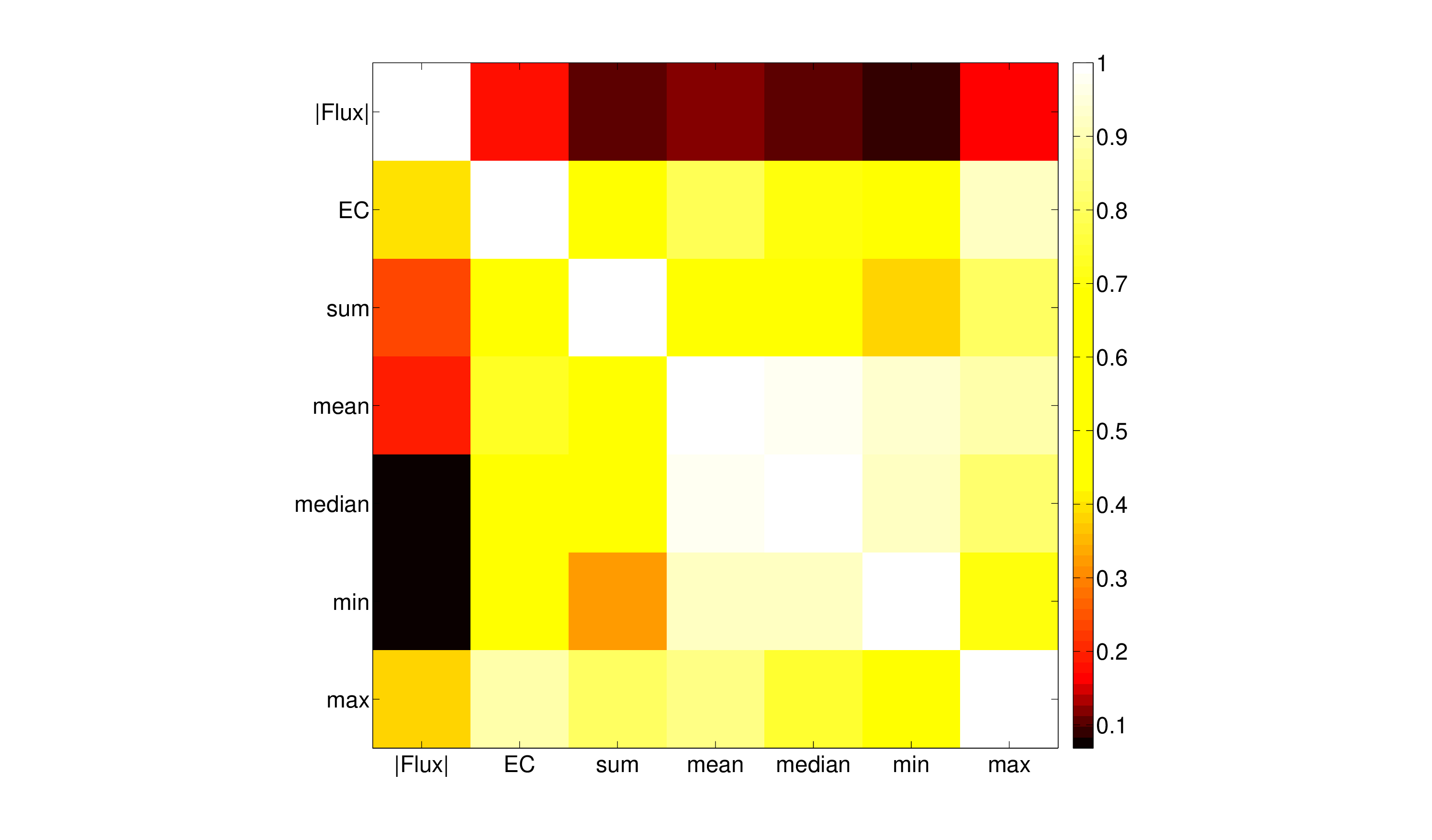}
  \caption{human} \label{fig:FluxExpCmp:B}
  \end{subfigure}
\\
\end{tabular}
\caption{Pearson correlation between FALCON flux magnitudes,
prerequisite enzyme complex estimates (from minDisj), and various
simpler gene expression estimates based on the list of genes
associated to each reaction. For yeast \textbf{(a)}, the upper and
lower triangles are the 75\% and 85\% maximum growth conditions,
respectively, and human is done similarly with the K562 and MDA-MB-231
cell lines \textbf{(b)}. As for expression estimates, the sum of
expression and enzyme complex estimate levels are generally the least
correlated with other expression estimates. As expected, the enzyme
complex estimates are the most correlated with the FALCON fluxes, as
they are used in the algorithm. However, it is important to note that
they are not very similar, exemplifying the affect the network
constraints play when determining flux. Interestingly, enzyme complex
abundance is found to correlate very highly with the maximum
expression level for the complex; this can be attributed to many genes
having relatively simple complexes that are isozymes, where one major
isozyme is typically highly expressed.}
\label{fig:FluxExpCmp}
\end{figure}
\FloatBarrier

\section{Assumptions for enzyme complex formation}
\label{sec:complexation}

In order to quantify enzyme complex formation (sometimes called enzyme
complexation), the notion of an enzyme complex should be formalized.
A protein complex typically refers to two or more physically
associated polypeptide chains, which is sometimes called a quaternary
structure. Since we are not exclusively dealing with multiprotein
complexes, we refer to an enzyme complex as being one or more
polypeptide chains that act together to carry out metabolic
catalysis.

\begin{figure}[!htb]
\begin{tabular}{c}
  \begin{subfigure}[b]{0.95\textwidth}
  \includegraphics[width=\textwidth]{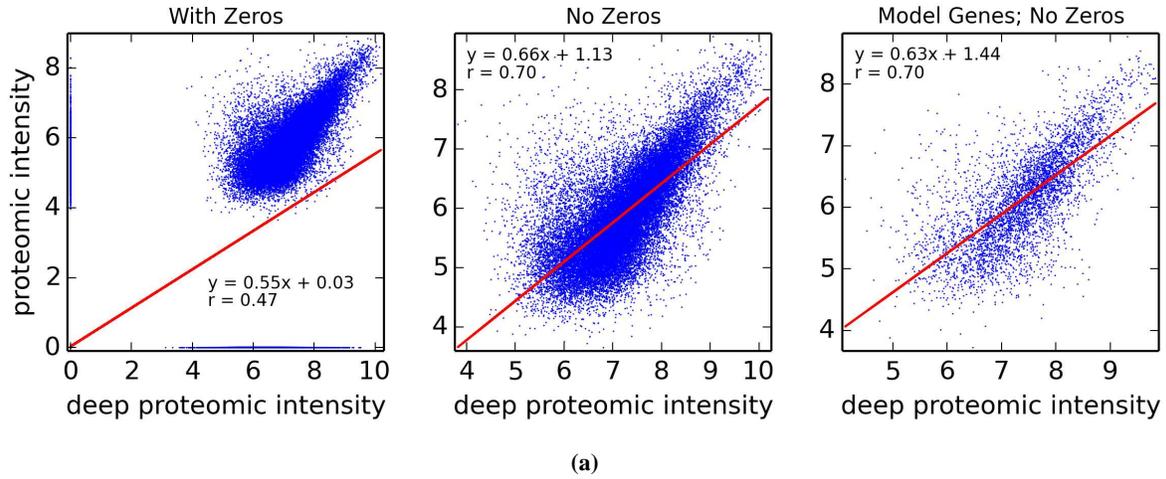}
  \caption{} \label{fig:HumanExpCorr:A}
  \end{subfigure}
\\
  \begin{subfigure}[b]{0.91\textwidth}
  \includegraphics[width=\textwidth]{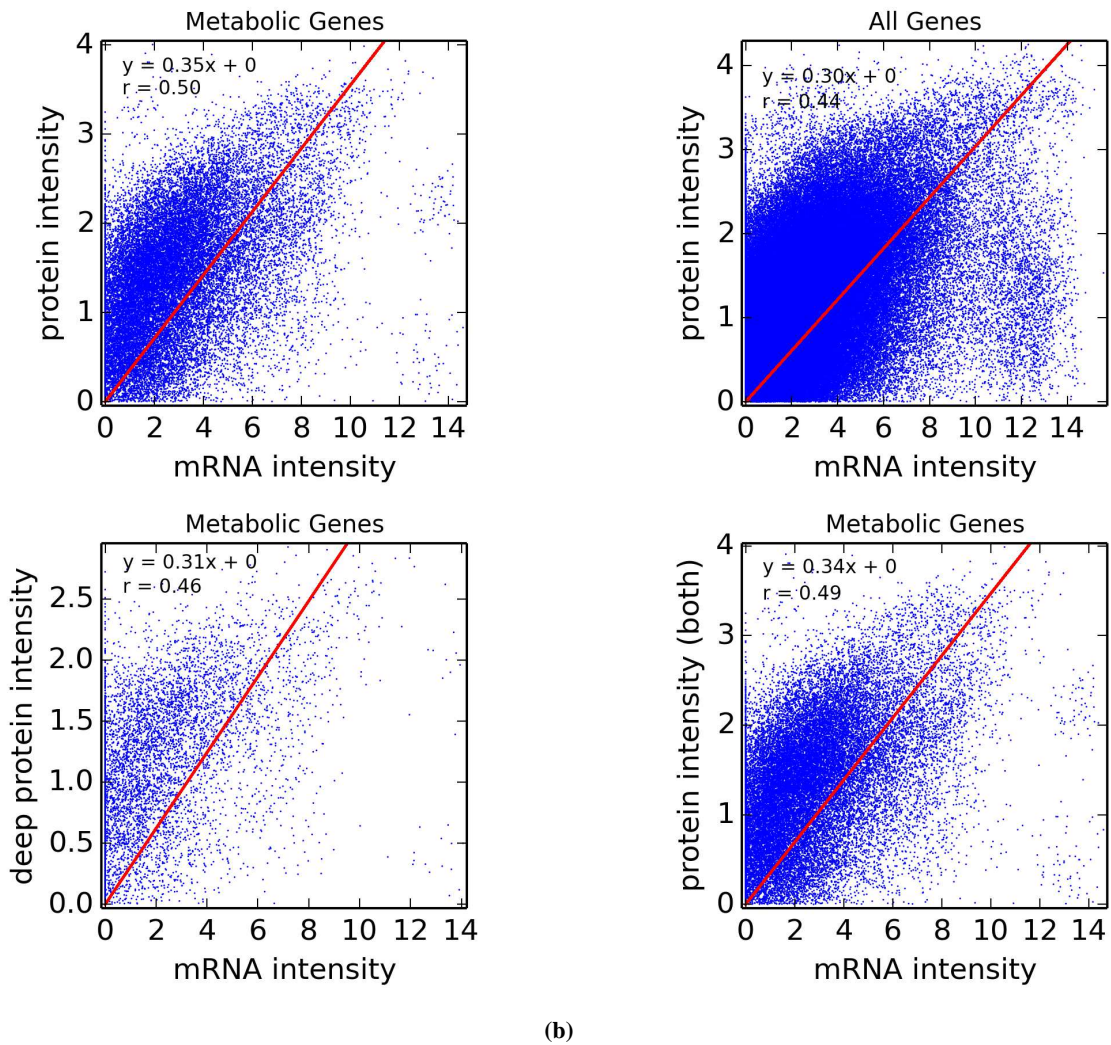}
  \caption{} \label{fig:HumanExpCorr:B}
  \end{subfigure} 
\\
\end{tabular}
\caption{Correlation between proteomic datasets obtained from two different
instruments \textbf{(a)}; 
left, proteins that were detected are shown as zeros, whereas they are filtered 
in the middle, and only metabolic genes are shown on the right. In \textbf{(b)},
these proteomic datasets (with zeros removed) are correlated with an 
mRNA (microarray) dataset.}
\label{fig:HumanExpCorr}
\end{figure}
\FloatBarrier

\emph{Assumption~\ref{asm:expcorr}.}  A fundamental assumption
that we need in order to guarantee an accurate estimate of (unitless)
enzyme complex abundance are the availability of accurate measurements
of their component subunits. Unfortunately, this is currently not
possible, and we almost always must make do with mRNA measurements,
which may even have some degree of inaccuracy in measuring the mRNA
abundance. What has been seen is that Spearman's $\rho = 0.6$ for
correlation between RNA-Seq and protein intensity in datasets from
HeLa cells \citep{Nagaraj2011}. We also found that there is moderate
correlation (Pearson's $r = 0.5$) between proteomic and microarray data
and fairly strong correlation between proteomic data obtained from
different instruments 
(Pearson's $r = 0.7$; \suppOrApp \Fig~\ref{fig:HumanExpCorr}).
This implies that much can likely
still be gleaned from analyzing various types of expression data, 
but, an appropriate
degree of caution must be used in interpreting results. 
By incorporating more information, such as metabolic
constraints, we hope to obviate some of the error in estimating
protein intensity from mRNA data.

\emph{Assumption~\ref{asm:isozyme}.} We also include the notion of
isozymes--different proteins that catalyze the same reaction--in our
notion of enzyme complex. Isozymes may arise by having one or more
differing protein isoforms, and even though these isoforms may not be
present in the same complex at the same moment, we consider them to be
part of the enzyme complex since one could be substituted for the
other.

\begin{figure*}
\centering
\includegraphics[clip=true,trim=0cm 0cm 0cm 0cm, width=12cm]{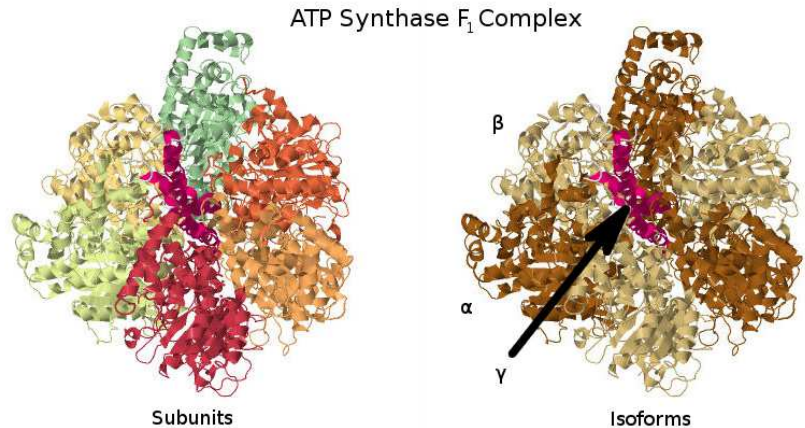}
\caption{Illustration of the $F_1$ part of the ATP Synthase complex
  (PDB ID 1E79; \protect\citealt{Gibbons2000,Bernstein1978,Gezelter}).
  This illustration demonstrates both how an enzyme complex may be
  constituted by multiple subunits (left), and how some of those
  subunits may be products of the same gene and have differing
  stoichiometries within the complex (right).}
\label{fig:2F43}
\end{figure*}

As an example for assumptions described so far, take the $F_1$
subcomplex of ATP Synthase (\Fig~\ref{fig:2F43}), which is composed
of seven protein subunits (distinguished by color, left). On the
right-hand side we see different isoforms depicted as different
colors.  Error in expression data aside, instead of considering the
abundances with multiplicity and dividing their expression values by
their multiplicity, it may be easier to simply note that the axle
peptide (shown in red in the center of the complex) only has one copy
in the complex, so its expression should be an overall good estimation
of the $F_1$ subcomplex abundance. This reasoning will be useful
later in considering why GPR rules may be largely adequate for estimating
the abundance of most enzyme complexes.

\emph{Assumption~\ref{asm:hierarchy}.}
The modeling of enzyme complex abundance can be tackled by using
nested sets of subcomplexes; each enzyme complex consists of multiple
subcomplexes, unless it is only a single protein or family of protein
isozymes.  These subcomplexes are required for the enzyme complex to
function (AND relationships), and can be thought of as the division of
the complex in to distinct units that each have some necessary
function for the complex, with the exception that we do not keep track
of the multiplicity of subcomplexes within a complex since this
information is, in the current state of affairs, not always known.
However, there may be alternative versions of each functional set
(given by OR relationships). Eventually, this nested embedding
terminates with a single protein or set of peptide isoforms
(e.g.\ isozymes).  In the case of ATP Synthase, one of its functional
sets is represented by the $F_1$ subcomplex. The $F_1$ subcomplex
itself can be viewed as having two immediate subcomplexes: the single
$\gamma$ (axle) subunit and three identical subcomplexes each made of
an $\alpha$ and $\beta$ subunit. Each $\alpha\beta$ pair works
together to bind ADP and catalyze the reaction \citep{Oster2003}. The
$\alpha\beta$ subcomplex itself then has two subcomplexes composed of
just an $\alpha$ subunit on the one hand and the $\beta$ subunit on
the other.  It is obvious that one of these base-level functional
subcomplexes (in this example, either $\gamma$ or $\alpha\beta$) will
be in most limited supply, and that it will best represent the overall
enzyme complex abundance (discounting the issues of multiplicity for
$\alpha\beta$, see Assumption~\ref{asm:nostoich}).

%
%

The hierarchical structure just described, when written out in
Boolean, will give a rule in CNF (conjunctive normal form), or more
specifically (owing to the lack of negations), clausal normal form,
where a clause is a disjunction of literals (genes). This is because all
relations are ANDs (conjunctions), except possibly at the inner-most
subcomplexes that have alternative isoforms, which are expressed as
ORs (disjunctions). Since GPR rules alone only specify the
requirements for enzyme complex formation, we will see that not all
forms of Boolean rules are equally useful in evaluating the enzyme
complex abundance, but we have established the assumptions in \suppOrApp
Table~\ref{tab:ECAssume} and an alternative and logically equivalent rule
\citep{Russell2009} under which we can estimate enzyme complex copy
number.

\begin{table}
\def \ECAssumeCap {Assumptions in GPR-based Enzyme Complex Formation}
  \begin{center} 
  \begin{tabular}{| p{0.9\textwidth} |}
  \hline
  \textbf{\suppOrApp Table~\ref{tab:ECAssume}. \ECAssumeCap} \\
  \hline
  \input{\commonDir falcon_assumptions}
  \\ \hline
  \end{tabular}
  \end{center}

\caption{A list of assumptions about how Gene-Protein-Reaction rules 
can describe enzyme complex stoichiometry.}
\label{tab:ECAssume}
\end{table}

There is no guarantee that a GPR rule has been written down with this
hierarchical structure in mind, though it is likely the case much of
the time as it is a natural way to model complexes.  However, any GPR
rule can be interpreted in the context of this hierarchical view due
to the existence of a logically equivalent CNF rule for any non-CNF
rule, and it is obvious that logical equivalence is all that is
required to check for enzyme complex formation when exact isoform
stoichiometry is unknown.  As an example, we consider another common
formulation for GPR rules, and a way to think about enzyme
structure---disjunctive normal form (DNF).  A DNF rule is a
disjunctive list of conjunctions of peptide isoforms, where each
conjunction is some variation of the enzyme complex due to
substituting in different isoforms for some of the required
subunits. A rule with a more complicated structure and compatible
isoforms across subcomplexes may be written more succinctly in CNF,
whereas a rule with only very few alternatives derived from isoform
variants may be represented clearly with DNF.  In rare cases, it is
possible that a GPR rule is written in neither DNF or CNF, perhaps
because neither of these two alternatives above are strictly the case,
and some other rule is more succinct.

\emph{Assumptions~\ref{asm:nostoich},~\ref{asm:sharing}~and~\ref{asm:active_site}.}
One active site per enzyme complex implies a single complex can only
catalyze one reaction at a time. Multimeric complexes with one active
site per identical subunit would be considered as one enzyme complex
per subunit in this model. Note that it is possible for an enzyme
complex to catalyze different reactions. In fact, some transporter
complexes can transfer many different metabolites across a lipid
bilayer---up to 294 distinct reactions in the reversible model for
solute carrier family 7 (Gene ID 9057).  Another example is the
ligation or hydrolysis of nucleotide, fatty acid, or peptide chains,
where chains of different length may all be substrates or products of
the same enzyme complex. While we do not explicitly consider these in
in the minimum disjunction algorithm, these redundancies are taken into
account subsequently in Algorithm~\ref{alg:FALCON}.

\begin{figure}[!htb]
\centering
\begin{tabular}{c}
  \begin{subfigure}[b]{0.75\textwidth}
  \includegraphics[width=\textwidth, height=0.4\textheight]{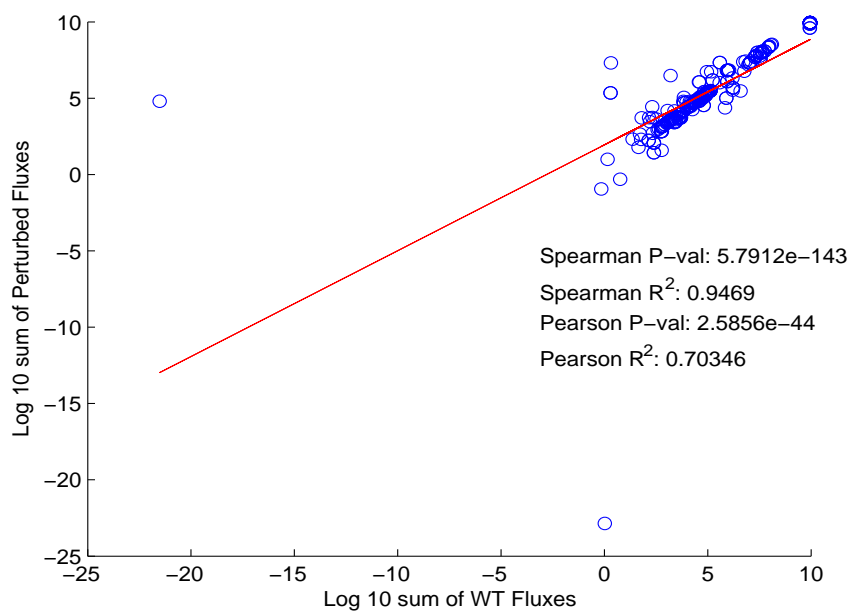}
  \caption{} \label{fig:WCComplexationFlux:A}
  \end{subfigure}
\\
  \begin{subfigure}[b]{0.75\textwidth}
  \includegraphics[width=\textwidth, height=0.4\textheight]{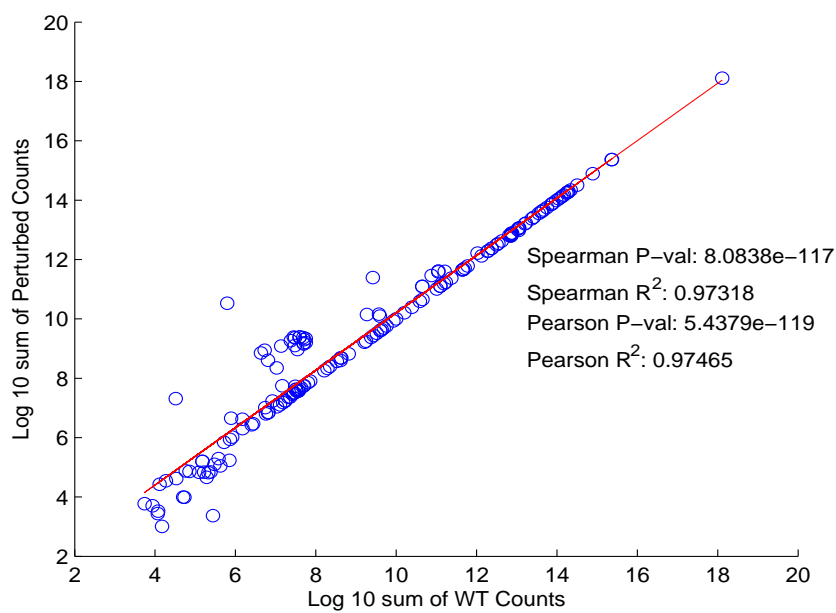}
  \caption{} \label{fig:WCComplexationFlux:B}
  \end{subfigure} 
\\
\end{tabular}
\caption{The Whole Cell model was simulated for 9000 seconds with both
default (WT) and simplified unitary (Perturb) protein complex
stoichiometry. Among 645 total metabolic fluxes, 298 had nonzero flux
in both conditions. We calculated the sum of each of these 298 fluxes
across 9000 seconds. We omitted 10 fluxes whose overall sum was of
different sign between WT and Perturb, took the $\log_{10}$ value of
the remaining 288 fluxes, and plotted these values, in both WT and
Perturb \textbf{(a)}. Similarly, among 722 total metabolites, 182 had
nonzero counts in both conditions. We calculated the sum of each of
the counts of these 182 metabolites across 9000 seconds, took the
$\log_{10}$ of these values, and plotted them in both WT and Perturb
\textbf{(b)}. In both cases, the red line indicates the line of best
fit for the Pearson's correlation.}
\label{fig:WCComplexationFlux}
\end{figure}

In order to explore the effect that stoichiometry of protein complexes
can have on metabolism, we compared time series aggregates of fluxes
and metabolite counts from a whole cell model of \emph{Mycoplasma
genitalium} \citep{Karr2012} in two different conditions (\suppOrApp
\Fig~\ref{fig:WCComplexationFlux}). While there are outliers and
deviations from perfect correlation, the results indicate that, 
at least in a whole cell model of \emph{M. genitalium}, our
assumption for protein complex stoichiometry may be acceptable.
Nonetheless, we hope to address this assumption at a future date.

What is currently not considered in our process is that some peptide
isoforms may find use in completely different complexes, and in some
cases, individual peptides may have multiple active sites; in the
first case, we assume an unrealistic case of superposition where the
isoform can simultaneously function in more than one complex. The
primary reason we have not tackled this problem is because exact
subunit stoichiometry of most enzyme complexes is not accurately
known, but an increasing abundance of data on BRENDA
\citep{Schomburg2013} gives some hope to this problem. A recent
\textit{E. coli} metabolic model incorporating the metabolism of all
known gene products \citep{O'Brien2013} also includes putative
enzyme complex stoichiometry in GPR rules. For the second point, there
are a few enzymes where a single polypeptide may have multiple active
sites (e.g.\ fatty acid synthase), and this is not currently taken into
account in our model. One way that we may take this into account has
been illustrated online by one of the authors \citep{Smallbone2014}. 

\FloatBarrier

\emph{Assumption~\ref{asm:holo}.}
We do not make any special assumptions requiring symmetry of an
isoform within a complex. For instance, the example in
assumption~\ref{asm:holo} shows how you might have one subcomponent
composed of a single isoform, and another subcomponent composed of
that gene in addition to another isoform. In this case, it is simply
reduced to being the first gene only that is required, since clearly
the second is strictly optional. That isn't to say that the second
gene may not have some metabolic effect, such as (potentially) aiding in
structural ability or altering the catalytic rate, but it should have
no bearing on the formation of a functional catalytic
complex. Holoenzymes---enzymes with metabolic cofactors or protein
subunits that have a regulatory function for the complex---would
likely be the only situation where this type of rule might need to be
considered in more detail. But in the absence of detailed kinetic
information, this consideration (much like allosteric
regulation) is not useful.

No additional algorithmic considerations are needed, as this is a
by-product of the conversion to CNF. For instance, take the following
example where the second conjunction has the redundant gene $g_3$: 

\[(g_1 \land g_2) \lor (g_1 \land g_2 \land g_3)\]

Distributing during the process of conversion to CNF results in:

\[
g_1 \land (g_1 \lor g_2 ) \land (g_1 \lor g_3) \land (g_2 \lor g_1) 
\land g_2 \land (g_2 \lor g_3)
\]

Because every disjunction with more than one literal is in conjunction
with another disjunction with only one of its literals, the
disjunction with fewer literals will be the minimum of the two once
evaluated. This applies to both of the singleton disjunctions $g_1$
and $g_2$, so all other disjunctions will effectively be ignored (it
is up to the implementer whether the redundant sub-expressions are
removed before evaluation):

\[
g_1 \land \bcancel{(g_1 \lor g_2 ) \land} \bcancel{(g_1 \lor g_3) \land} 
\bcancel{(g_2 \lor g_1) \land} g_2 \bcancel{\land (g_2 \lor g_3)} 
= (g_1 \land g_2)
\]

\emph{Assumption~\ref{asm:enzyme_sensitivity}.}
Another important biochemical assumption is that reactions should
operate in a regime where they are sensitive to changes in the overall
enzyme level in the pathways that they belong in
\citep{Bennett2009,Chubukov2013}. This is perhaps the most important
issue to be explored further for methods like this, since if it is not
true, some other adjustment factor would be needed to make the method
realistic. For instance, if all reactions in a pathway are operating
far below $V_{max}$, but it is not the case in another pathway, the
current method does not have information on this, and will try to put
more flux through the first pathway than should be the case.

\emph{Assumptions~\ref{asm:chap},~\ref{asm:posttrans}~and~\ref{asm:rate}.}
Due to the quickness, stability, and energetic favorability of enzyme
complex formation, the absence of chaperones or coupled metabolic
reactions required for complex formation may be reasonable
assumptions, but further research is warranted \citep{Karr2012}.
Additionally, as in metabolism, we assume a steady state for complex
formation, so that rate laws regarding complex formation aren't
needed. However, further research may be warranted to investigate the
use of a penalty for complex levels based on mass action and
protein-docking information. Requisite to this would be addressing
assumption~\ref{asm:nostoich}. It would be surprising (but not
impossible) if such a penalty were very large due to the cost this
would imply for many of the large and important enzyme complexes
present in all organisms \citep{Nelson2008}. A more serious
consideration may be that information on post-translational
modification is not currently considered. Post-translational
modification is highly context-specific and the relevant data is not
as cheap to get as expression data, so it may be some time before it
can be integrated into the modeling framework.

\clearpage

\section{Benchmarking of solvers}
We have exclusively used the Gurobi solver \citep{gurobi} for this
work, which is a highly competitive solver that employs by default a
parallel strategy to solving problems: a different algorithm is run
simultaneously, and as soon as one algorithm finished the others
terminate. Of course, if there is a clear choice of algorithm for a
particular problem class, this should be used in production settings
to avoid wasted CPU time and memory. In order to address this, we
benchmarked the three non-parallel solver methods in Gurobi
 (since parallel solvers simply use multiple methods simultaneously).
The exception to this rule is the Barrier method, which can use
multiple threads, but in practice for our models appears to use
no more than about 6 full CPU cores simultaneously for our models.
Our results for Yeast~5 and Yeast~7 with minimal directionality constraints
\citep{Heavner2012,Lee2012,Aung2013} and Human Recon~2 \citep{Thiele2013}
are shown in \suppOrApp Table~\ref{tab:methodTime}).

\begin{table}
\begin{center}
\begin{tabular}{rrrr}
\emph{Model}                 & \emph{Primal-Simplex} & \emph{Dual-Simplex} & \emph{Barrier} \\
Yeast~5.21 (2,061 reactions) & $ 7.841 \pm 1.697    $ & $ 7.611 \pm 1.267    $ & $ 10.859 \pm 2.788   $\\ 
Yeast~7.0 (3,498 reactions)  & $ 51.863 \pm 22.731  $ & $ 65.317 \pm 12.771  $ & $ 242.137 \pm 57.129 $\\
Human 2.03 (7,440 reactions) & $ 159.077 \pm 24.903 $ & $ 152.297 \pm 39.783 $ & $ 366.166 \pm 92.321 $\\
\end{tabular}
\end{center}
\caption{Running times (in seconds, $\pm$ standard deviation) for
  FALCON using various algorithms implemented in the Gurobi package.
  For yeast models, 1,000 replicates were performed, and for the human
  model, 100 replicates were performed.}
\label{tab:methodTime}
\end{table}

We found that in Yeast~7 with the primal-simplex solver, there is a
chance the solver will fail to find a feasible solution.
We verified that this is a numeric issue
in Gurobi and can be fixed by setting the Gurobi parameter
\texttt{MarkowitzTol} to a larger value (which decreases
time-efficiency but limits the numerical error in the
simplex algorithm). In practice, failure for the algorithm to converge
at an advanced iteration is rare and is not always a major problem (since the previous
flux estimate by the advanced iteration should already be quite good), but it
is certainly undesirable; a warning message will be printed by
\texttt{falcon} if this occurs, at which point parameter settings can
be investigated. In the future, we plan to improve \texttt{falcon} so
that parameters will be adjusted as needed during progression of the
algorithm after finding a good test suite of models and data. For now,
we use the dual-simplex solver, for which we have always had good
results.

Because the number of iterations depends non-trivially on the model
and the expression data, it may be more helpful to look at the 
average time per iteration in the above examples 
(\suppOrApp Table~\ref{tab:methodTimeIter}).

\begin{table}
\begin{center}
\begin{tabular}{rrrr}
\emph{Model}                 & \emph{Primal-Simplex} & \emph{Dual-Simplex} & \emph{Barrier} \\
Yeast~5.21 (2,061 reactions) & $ 0.721 \pm 0.023 $ & $ 0.652 \pm 0.040 $ & $ 1.100 \pm 0.112  $\\ 
Yeast~7.0 (3,498 reactions)  & $ 2.725 \pm 0.298 $ & $ 2.469 \pm 0.289 $ & $ 11.309 \pm 1.589 $\\
Human 2.03 (7,440 reactions) & $ 6.422 \pm 0.484 $ & $ 5.233 \pm 0.661 $ & $ 15.782 \pm 3.209 $\\ 
\end{tabular}
\end{center}
\caption{Running time per FALCON iteration (in seconds, $\pm$ standard
  deviation) using various algorithms implemented in the Gurobi
  package.  For yeast models, 1,000 replicates were performed, and for
  the human model, 100 replicates were performed.}
\label{tab:methodTimeIter}
\end{table}

Given the above rare trouble with primal simplex solver the universal
best performance enjoyed by the dual-simplex method (\suppOrApp Tables
\ref{tab:methodTime} and \ref{tab:methodTimeIter}), we would advise
the dual-simplex algorithms, all else being equal. The dual-simplex
method is also recommended for memory-efficiency by Gurobi
documentation, but we did not observe any differences in memory for
different solver methods.

All timing analyses were performed on a system with four 8-core AMD
Opteron\texttrademark\ 6136 processors operating at 2.4
GHz. \Fig~\ref{fig:FluxBars}, Table~\ref{tab:FalcPerf}, and \suppOrApp
Tables \ref{tab:methodTime} and \ref{tab:methodTimeIter} used a single
unperturbed expression file per species (\textit{S. cerevisiae} and
\textit{H. sapiens}; see \texttt{timingAnalysis.m} for
details). Values were averaged across 32 replicates. \suppOrApp
Note that the iMAT method is formulated as a mixed integer program
\citep{Shlomi2008}, and was able to use additional parallelization of
the solver \citep{gurobi} whereas other methods only used a single
core (our system had 32 cores and iMAT with Gurobi would use all of
them). Tables~\ref{tab:methodTime} and \ref{tab:methodTimeIter} used
multivariate log-normal noise multiplied by the original expression
vector to introduce more variance in the calculations; the human
models were tested with 100 replicates and the yeast models with 500
replicates.

\section{Generation of figures and tables}

All non-trivial figures can be generated using MATLAB scripts found in
the \texttt{analysis/figures} subdirectory of the FALCON installation.
In particular, figures should be generated through the master script
\texttt{makeMethodFigures.m} by calling
\texttt{makeMethodFigures(figName)} where \texttt{figName} has a name
corresponding to the desired figure.  In some cases, some MATLAB
\texttt{.mat} files will need to be generated by other scripts first;
see the plotting scripts or the subsections below for details. An
example is to make the scatter plots showing the difference between
running falcon with enzyme abundances determined by direct evaluation
or the minimum disjunction algorithm; all three scatter plots are
generated with the command \texttt{makeMethodFigures(\textquotesingle
fluxCmpScatter\textquotesingle)}. Note that, as written, this requires
a graphical MATLAB session.

Several figures can be generated that are related to comparing 
human proteomic and transcriptomic data by executing the script 
\texttt{proteomic\_file\_make.py} in the \texttt{analysis/nci60}
subdirectory; this includes \suppOrApp \Fig~\ref{fig:HumanExpCorr}.

\begin{sloppypar}
Comparison of the effects of the employed enzyme complexation methods
were evaluated using \texttt{compareEnzymeExpression.m} and 
\texttt{compareFluxByRGroup.m}. Comparison of reaction groups was
performed in \texttt{compareFluxByRGroup.m}.
\end{sloppypar}

\subsection{Timing analyses}
All timing analyses were performed on a system with four 8-core AMD
Opteron\texttrademark\ 6136 processors operating at 2.4 GHz. 
\Fig~\ref{fig:FluxBars} and Table~\ref{tab:FalcPerf} used unperturbed
expression data; see
\texttt{yeastResults.m} for details). Values for the FALCON method
were averaged across 32 replicates, while values for the
\citealt{Lee2012} method were averaged across 8 replicates. Human
timing analyses were performed using \texttt{methodTimer.m} with
8 replicates.

\subsection{Data sources}
Enzyme complexation comparisons were performed on proteomics data
from \citealt{Gholami2013} (Human; 786-O cell line) and 
\citealt{Picotti2013} (yeast; BY strain), and on RNA-Seq data
from \citealt{Lee2012} (yeast; 75\% max $\mu$ condition).
Human proteomic and trascriptomic data used for comparison
in \suppOrApp \Fig~\ref{fig:HumanExpCorr} were taken from 
\citealt{Gholami2013}.

%% file: common/falcon_assumptions.tex


\begin{enumerate}
\item \label{asm:expcorr}
Expression values are highly correlated with the copy numbers of their
corresponding peptide isoforms.
\item \label{asm:isozyme} 
Protein isoforms contributing to isozymes are considered part of the
same enzyme complex.
\item \label{asm:hierarchy}
Any enzyme complex can be described as a hierarchical subset of
(possibly redundant) subcomplexes; redundant subcomplexes, as
elaborated in (\ref{asm:nostoich}), are not currently modeled.
\item \label{asm:nostoich} 
Assume one copy of peptide per complex; exact isoform stoichiometry
is not considered.
\item \label{asm:sharing} 
With the exception of complexes having identical rules (i.e. the same
complex listed for different reactions), each copy of a peptide
is available for all complexes in the model.
\item \label{asm:active_site}
There is only one active site per enzyme complex.
\item \label{asm:enzyme_sensitivity} 
We assume that different pathways have similar flux sensitivities
with respect to their enzyme abundances.
\item \label{asm:holo} 
If a particular subcomplex can be catalyzed by A and it can also be
catalyzed by A and B (e.g. B acts as a regulatory unit, as in
holoenzymes), this just simplifies to A once expression values are
substituted in. Similarly, allosteric regulation is not
modeled. Relatedly, there are no NOT operations in GPR rules (just ANDs
and ORs).
\item \label{asm:chap} 
Enzyme complexes form without the assistance of protein chaperones and
formation is not coupled to other reactions.
\item \label{asm:posttrans}
Post-translational modifications do not affect complex formation.
\item \label{asm:rate} 
Rate of formation and degradation of complexes doesn't play a role,
since we assume steady-state. 
\end{enumerate}